\documentclass[twoside,twocolumn,9pt]{article}
\usepackage{extsizes}
\usepackage[super,sort&compress,comma]{natbib} 
\usepackage[version=3]{mhchem}
\usepackage[left=1.5cm, right=1.5cm, top=1.785cm, bottom=2.0cm]{geometry}
\usepackage{balance}
\usepackage{times,mathptmx}
\usepackage{sectsty}
\usepackage{graphicx} 
\usepackage{lastpage}
\usepackage[format=plain,justification=justified,singlelinecheck=false,font={stretch=1.125,small,sf},labelfont=bf,labelsep=space]{caption}
\usepackage{float}
\usepackage{fancyhdr}
\usepackage{fnpos}
\usepackage[english]{babel}
\addto{\captionsenglish}{%
  
}
\usepackage{array}
\usepackage{droidsans}
\usepackage{charter}
\usepackage[T1]{fontenc}
\usepackage[usenames,dvipsnames]{xcolor}
\usepackage{setspace}
\usepackage[compact]{titlesec}
\usepackage{hyperref}

\usepackage{amssymb}
\usepackage{gensymb}
\usepackage{color}
\usepackage{siunitx}
\usepackage{upgreek}

\newcommand*{\citen}[1]{%
  \begingroup
    \romannumeral-`\x 
    \setcitestyle{numbers}%
    \cite{#1}%
  \endgroup   
}

\usepackage{epstopdf}

\definecolor{cream}{RGB}{222,217,201}

\begin{document}

\pagestyle{fancy}
\thispagestyle{plain}
%

\makeFNbottom
\makeatletter
\renewcommand\LARGE{\@setfontsize\LARGE{15pt}{17}}
\renewcommand\Large{\@setfontsize\Large{12pt}{14}}
\renewcommand\large{\@setfontsize\large{10pt}{12}}
\renewcommand\footnotesize{\@setfontsize\footnotesize{7pt}{10}}
\makeatother

\renewcommand{\thefootnote}{\fnsymbol{footnote}}
\renewcommand\footnoterule{\vspace*{1pt}%
\color{cream}\hrule width 3.5in height 0.4pt \color{black}\vspace*{5pt}} 
\setcounter{secnumdepth}{5}

\makeatletter 
\renewcommand\@biblabel[1]{#1}            
\renewcommand\@makefntext[1]%
{\noindent\makebox[0pt][r]{\@thefnmark\,}#1}
\makeatother 
\renewcommand{\figurename}{\small{Fig.}~}
\sectionfont{\sffamily\Large}
\subsectionfont{\normalsize}
\subsubsectionfont{\bf}
\setstretch{1.125} 
\setlength{\skip\footins}{0.8cm}
\setlength{\footnotesep}{0.25cm}
\setlength{\jot}{10pt}
\titlespacing*{\section}{0pt}{4pt}{4pt}
\titlespacing*{\subsection}{0pt}{15pt}{1pt}


\makeatletter 
\newlength{\figrulesep} 
\setlength{\figrulesep}{0.5\textfloatsep} 

\newcommand{\topfigrule}{\vspace*{-1pt}%
\noindent{\color{cream}\rule[-\figrulesep]{\columnwidth}{1.5pt}} }

\newcommand{\botfigrule}{\vspace*{-2pt}%
\noindent{\color{cream}\rule[\figrulesep]{\columnwidth}{1.5pt}} }

\newcommand{\dblfigrule}{\vspace*{-1pt}%
\noindent{\color{cream}\rule[-\figrulesep]{\textwidth}{1.5pt}} }

\makeatother

\twocolumn[
  \begin{@twocolumnfalse}

\noindent\LARGE{\textbf{Weak carbohydrate-carbohydrate interactions in membrane adhesion are fuzzy and generic}}
\\ 


 \noindent\large{Batuhan Kav,\textit{$^{a}$}  Andrea Grafm\"uller,\textit{$^{a}$} Emanuel Schneck,\textit{$^{b,c}$}   and  Thomas R.\ Weikl\textit{$^{a}$}} \\

\noindent\normalsize{Carbohydrates such as the trisaccharide motif Le$^{\rm X}$ are key constituents of cell surfaces. Despite intense research, the interactions between carbohydrates of apposing cells or membranes are not well understood. In this article, we investigate carbohydrate-carbohydrate interactions in membrane adhesion as well as in solution with extensive atomistic molecular dynamics simulations that exceed the simulation times of previous studies by orders of magnitude. For Le$^{\rm X}$,  we obtain association constants of soluble carbohydrates, adhesion energies of lipid-anchored carbohydrates, and maximally sustained forces of carbohydrate complexes in membrane adhesion that are in good agreement with experimental results in the literature. Our simulations thus appear to provide a realistic, detailed picture of Le$^{\rm X}$--Le$^{\rm X}$ interactions in solution and during membrane adhesion. In this picture, the Le$^{\rm X}$--Le$^{\rm X}$ interactions are  fuzzy, i.e.~Le$^{\rm X}$ pairs interact in a large variety of short-lived, bound conformations. For the synthetic tetrasaccharide Lac 2, which is composed of two lactose units, we observe similarly fuzzy interactions and obtain association constants of both soluble and lipid-anchored variants  that are comparable to the corresponding association constants of Le$^{\rm X}$. The fuzzy, weak carbohydrate-carbohydrate interactions quantified in our simulations thus appear to be a generic feature of small, neutral carbohydrates such as Le$^{\rm X}$ and Lac 2.
}


 \end{@twocolumnfalse} \vspace{0.6cm}

  ]

\renewcommand*\rmdefault{bch}\normalfont\upshape
\rmfamily
\section*{}
\vspace{-1cm}


\footnotetext{\textit{$^{a}$~Max Planck Institute of Colloids and Interfaces, Department of Theory and Bio-Systems, Am M\"uhlenberg 1, 14476 Potsdam, Germany.}}
\footnotetext{\textit{$^{b}$~Max Planck Institute of Colloids and Interfaces, Department of Biomaterials, Am M\"uhlenberg 1, 14476 Potsdam, Germany.}}
\footnotetext{\textit{$^{b}$~Technische Universit\"at Darmstadt, Physics Department, Hochschulstraße 12, 64289 Darmstadt, Germany.}}





\section*{Introduction}

Carbohydrates are omnipresent at cell surfaces as constituents of glycolipids and glycoproteins \cite{Alberts14,Varki07,Dennis09}. During cell adhesion, these carbohydrates get in touch with proteins and carbohydrates on apposing cell surfaces. While specific interactions between carbohydrates and proteins are known to play important roles in cell adhesion events, the role of carbohydrate-carbohydrate interactions in these events is less clear \cite{Schnaar04,Liu05,Arnaud13,Varki17,Poole18}. About three decades ago, homophilic carbohydrate-carbohydrate interactions of the trisaccharide Lewis$^{\rm X}$  (Le$^{\rm X}$) have been reported to be involved in embryonal cell compaction and aggregation \cite{Eggens89,Kojima94,Handa07}, and interactions between long carbohydrate chains have been linked to  the species-specific aggregation of marine sponges \cite{Misevic87}. In the following decades, carbohydrate-carbohydrate interactions in adhesion have been investigated in a variety of reconstituted or synthetic systems including nanoparticles and surfaces functionalized with carbohydrates \cite{Fuente01,Hernaiz02,Fuente05}, atomic force microscopy setups \cite{Tromas01,Bucior04,Lorenz12,Witt16},  and reconstituted vesicles \cite{Gourier05,Kunze13} or membranes \cite{Yu98,Schneck11,Latza20} containing glycolipids. While some carbohydrate-carbohydrate interactions have been reported to be strong \cite{Bucior04,Day15,Yu19}, interactions of small, neutral carbohydrates are typically considered to be weak \cite{Pincet01,Patel07}. However, the binding association constants, in particular at membrane interfaces, and the structural binding mechanisms are often not known.  

\begin{figure*}
\centering
\includegraphics[width=0.9\linewidth]{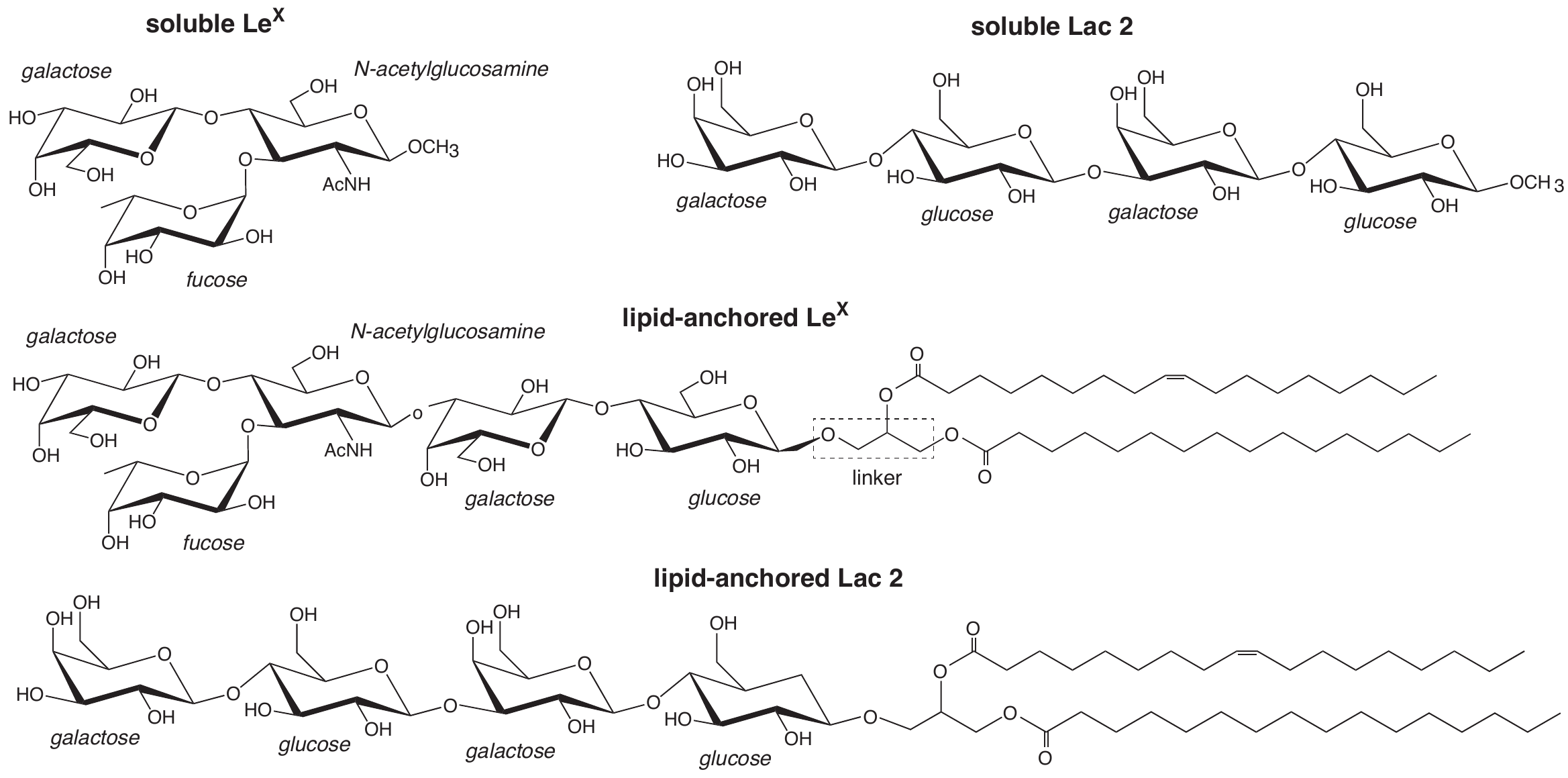}
\caption{Structures of the soluble and lipid-anchored saccharides investigated in our simulations.}
\label{figure-structure}
\end{figure*}
\begin{figure*}[t]
\includegraphics[width=\linewidth]{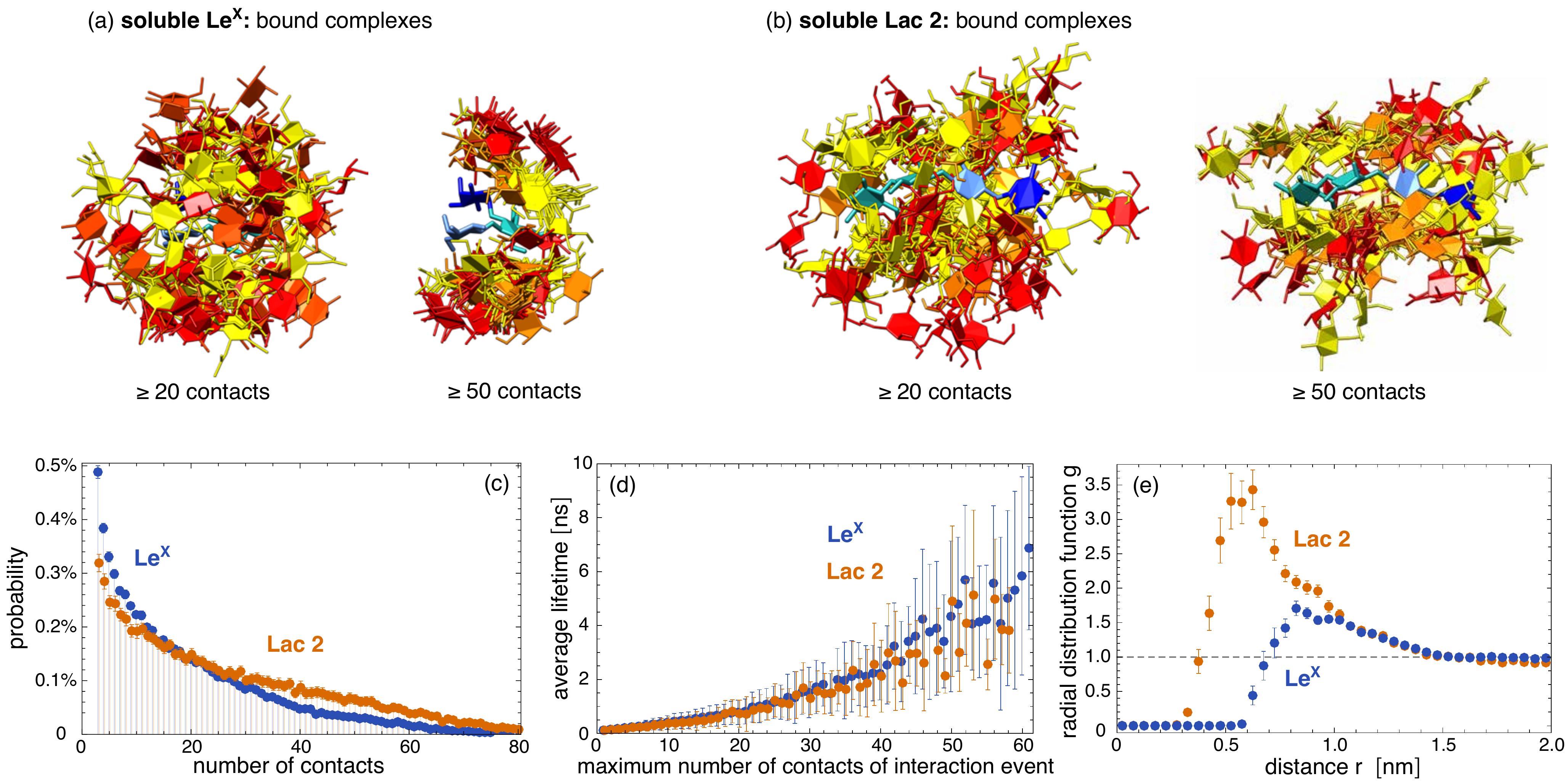}
\caption{(a) and (b) Randomly selected pair conformations of two Le$^{\rm X}$ and two Lac 2 molecules with at least 20 or at least 50 contacts between non-hydrogen atoms within a distance less than $0.45$ nm, respectively. One of the molecules is aligned in the 50 pair conformations and represented in blue colors, while the other molecule is represented in red/yellow colors. In the aligned Le$^{\rm X}$ molecules, fucose is represented in dark blue, galactose in light blue, and N-acytylglucosamine in cyan. In the other Le$^{\rm X}$ molecules, these monosaccharide units are represented in red, orange, and yellow, respectively.  In the aligned Lac 2 molecules, the terminal galactose is represented in dark blue, the adjacent glucose in light blue, and the remaining galactose and glucose in cyan. In the other Lac 2 molecules, these monosaccharides are shown in red, orange, and yellow. (c) Probability distributions of the number of contacts between non-hydrogen atoms obtained from our simulations of two soluble Le$^{\rm X}$ or two soluble Lac 2 molecules. (d) Average lifetime of interaction events as a function of the maximum number of contacts of the interaction events. Interaction events are consecutive stretches of simulation frames at intervals of $0.1$ ns with nonzero contacts of the two molecules. The error bars represent the standard deviations of the observed lifetimes. (e) Radial distribution functions $g(r)$ of two soluble Le$^{\rm X}$ or Lac 2 molecules with center-of-mass distance $r$. 
}
\label{figure-soluble}
\end{figure*}

In this article, we present detailed results from atomistic molecular dynamics simulations of carbohydrate-carbohydrate  interactions in membrane adhesion and in solution for Le$^{\rm X}$ and the synthetic saccharide Lac 2, which is composed of two lactose units \cite{Schneider01} (see Fig.\ \ref{figure-structure}). Our simulations employ a recent carbohydrate force field \cite{Sauter16} that allows a more faithful representation of carbohydrate-carbohydrate interactions \cite{Sauter16,Sauter16,Lay16,Woods18} and exceed the times and system sizes in previous simulation studies of carbohydrate-carbohydrate interactions in solution \cite{Gourmala07,Luo08,Santos09} by orders of magnitude.  Le$^{\rm X}$ has been investigated extensively as a model system for carbohydrate-carbohydrate interactions \cite{Fuente01,Tromas01,Hernaiz02,Gourier05,Schneck11,Kunze13,Witt16}, and  experimental data available from these investigations are central to corroborate our simulation results. In our  Le$^{\rm X}$ glycolipids, the Le$^{\rm X}$ trisaccharide is connected {\em via} a lactose disaccharide and a glycerol linker to lipids tails (see Fig.\ \ref{figure-structure}).  In our Lac 2 glycolipids, the  Le$^{\rm X}$ trisaccharide  is replaced by another lactose disaccharide, which allows to compare the carbohydrate-carbohydrate interactions of Le$^{\rm X}$ to those of the common saccharide lactose. From simulations of soluble pairs of Le$^{\rm X}$ and Lac 2, we obtain association constants $K_a$ of the order of $10$ M$^{-1}$, which agrees with a $K_a$ value of  Le$^{\rm X}$ derived from weak affinity chromatography experiments \cite{Fuente04,Tromas01}. From simulations of pairs of Le$^{\rm X}$ and Lac 2 glycolipids at apposing membrane surfaces, we obtain comparable association constants $K_{\rm trans}$ for the Le$^{\rm X}$ and Lac 2  glycolipids that strongly decrease with increasing membrane separation. For the membrane separation and thermal roughness of membrane multilayers with $10$ mol\% Le$^{\rm X}$ glycolipids measured in neutron scattering experiments \cite{Schneck11}, we determine an adhesion energy per area of the order of $10$ $\mu$J/m from our $K_{\rm trans}$ values, in agreement with the adhesion energy per area reported for vesicles that contain $10$ mol\% of Le$^{\rm X}$ glycolipids \cite{Gourier05}. The average force on bound Le$^{\rm X}$ glycolipid complexes determined in our simulations increases with increasing membrane separation up to a maximum value of about $20$ pN, which agrees with the Le$^{\rm X}$--Le$^{\rm X}$  unbinding force obtained from atomic force microscopy experiments \cite{Tromas01}. The agreement with experimental results indicates that our simulations provide a realistic, detailed picture of weak carbohydrate-carbohydrate interactions in solution as well as in membrane adhesion. A striking feature is that the carbohydrate-carbohydrate interactions are fuzzy, i.e.~both soluble and lipid-anchored variants of Le$^{\rm X}$ and Lac 2 interact in our simulations {\em via} a large variety of diverse, bound conformations.  

\section*{Results}

\subsection*{Interactions of soluble carbohydrates}

We first consider the interaction of two Le$^{\rm X}$ trisaccharides in solution and compare this Le$^{\rm X}$--Le$^{\rm X}$ pair interaction to the interaction of two Lac 2 tetrasaccharides, which are composed of two lactose units \cite{Schneider01} (see Fig.\ \ref{figure-structure}). Standard carbohydrate force fields lead to osmotic pressures for solutions of neutral carbohydrates that are systematically too low compared to experimental values.  This underestimation of the osmotic pressure of the carbohydrate solutions results from an overestimation of attractive carbohydrate-carbohydrate interactions \cite{Sauter16,Lay16}. To avoid unrealistically attractive carbohydrate-carbohydrate interactions, we have used the GLYCAM06$^{\rm TIP5P}_{\rm OSMOr14}$ force field, in which the van der Waals parameters for saccharide-saccharide interactions of the standard force field GLYCAM06 have been reparametrized to correctly reproduce experimentally measured osmotic pressures \cite{Sauter16}. The GLYCAM06$^{\rm TIP5P}_{\rm OSMOr14}$ force field employs the TIP5P water model because this water model leads to more reliable carbohydrate-carbohydrate interactions in GLYCAM06, compared to the standard TIP3P water model \cite{Sauter15,Woods18}. Using graphics processing units (GPUs) and the software AMBER GPU \cite{Salomon-Ferrer13,Le-Grand13}, we have generated $50$ simulation trajectories with a length of $2.0$ $\mu$s for two Le$^{\rm X}$ molecules in a periodic simulation box of volume $V = 131.5\, {\rm nm}^3$, and 40 trajectories with a length of $1$  $\mu$s or close to $1$  $\mu$s for two Lac 2 molecules in a simulation box of volume $V = 260.5\, {\rm nm}^3$, at the simulation temperature 30$\degree$C. Our total simulation times are $100$ $\mu$s for the Le$^{\rm X}$ pair and 39.5 $\mu$s for the Lac 2 pair, which greatly exceed the total simulation times up to $40$ ns \cite{Luo08} in previous simulation studies of Le$^{\rm X}$-Le$^{\rm X}$ pair interactions in solution \cite{Gourmala07,Luo08} and the total simulation time of a few ns for pair interactions of trisaccharide epitopes from marine sponges \cite{Santos09}.

In our simulations, we observe thousands of interaction events in which the two Le$^{\rm X}$ molecules or the two Lac 2 molecules are in contact. These interaction events are separated by longer or shorter trajectory parts in which the two molecules are not in contact. Figs.\ \ref{figure-soluble}(a) and (b) display pair conformations of Le$^{\rm X}$ and Lac 2  in which the two molecules exhibit at least 20 or 50 contacts of non-hydrogen atoms, respectively.  The shown pair conformations are randomly selected from the simulation frames of our trajectories. One of the carbohydrate molecules is aligned in the pair conformations and represented in blue colors, while the other molecule is represented in red/yellow colors. The clouds of red/yellow molecules around the aligned blue molecules in these conformations illustrate that the carbohydrate-carbohydrate interactions are `fuzzy' \cite{Tompa08,Uversky12}, i.e.~the two molecules interact in broad ensembles of conformations, rather than {\em via} a single binding conformation. For both Le$^{\rm X}$ and Lac 2, the ensembles of conformations with at least 50 contacts are narrower than the ensembles of conformations with at least 20 contacts. In conformations with at least 50 contacts, the two Le$^{\rm X}$ molecules tend to stack above each other in different orientations, and the two Lac 2 molecules tend to align parallel or anti-parallel. However, the probability distributions of contact numbers in Fig.\ \ref{figure-soluble}(c) illustrate that pair conformations with 50 or more contacts of non-hydrogen atoms are rather rare and  not typical. The probability distributions decrease monotonously with increasing number of contacts. 

\begin{table}[b]
  \caption{Association constants $K_a$ in units of M$^{-1}$ for different cutoffs $n_c$ for the contact number of binding events}
  \label{tbl:notes}
  \begin{tabular}{c|ccc}
  &   $n_c = 5$     &     $n_c = 10$      &     $n_c = 20$    \\
    \hline \\[-0.3 cm]
 Le$^{\rm X}$   & $6.4\pm 0.3$  &   $5.7\pm 0.3$  &  $4.5\pm 0.3$  \\
 Lac 2 & $13.2\pm 1.0$ & $12.3\pm 1.0$ & $10.7\pm 0.9$ \\
  \end{tabular}
\end{table}
\begin{figure*}
\includegraphics[width=\linewidth]{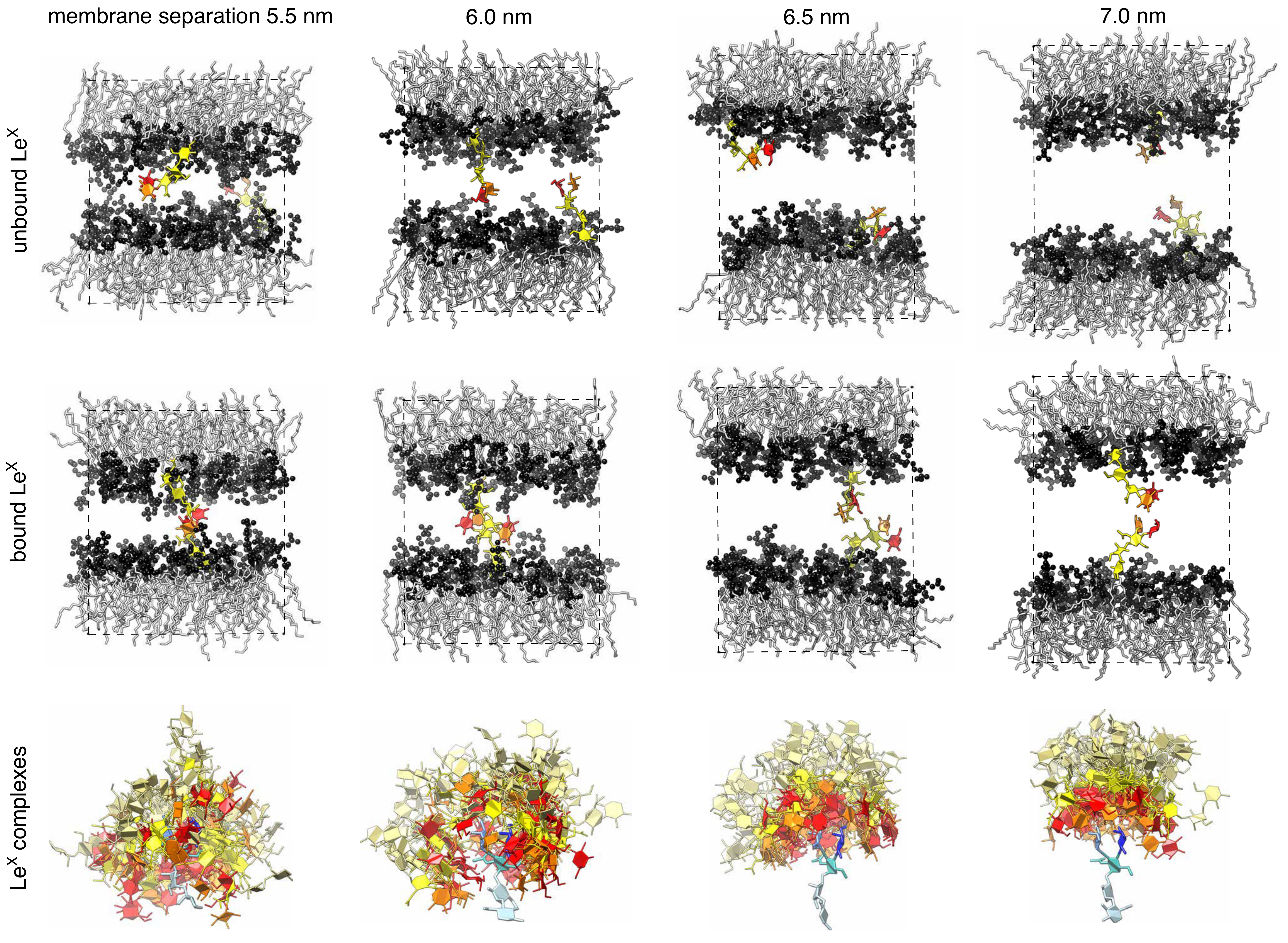}
\caption{(top) Membrane conformations with two unbound or bound Le$^{\rm X}$ glycolipids. The Le$^{\rm X}$ glycolipids are anchored in the different monolayers of the membrane and interact because of the periodic boundary conditions of the simulation box. The height of the simulation box corresponds to the membrane separation from bilayer midplane to midplane. Each membrane monolayer contains 35 POPC lipids, which have the same lipid tails as the Le$^{\rm X}$ glycolipids. The fucose and galactose at the branched tip of the Le$^{\rm X}$ glycolipids are represented in red and orange, and the remaining three monosaccharide units in yellow. (bottom) 50 randomly selected complexes of the carboyhydrate tips of the Le$^{\rm X}$ glycolipids  at different membrane separations. The selected complexes exhibit at least 10 contacts between non-hydrogen atoms of the two carbohydate tips. The carbohydrate tip of the lower Le$^{\rm X}$ glycolipid is aligned in the 50 complexes and represented in blue colors, while the carbohydrate tip of the upper glycolipid is represented in red/yellow colors. The Le$^{\rm X}$ motif of the carbohydrate tips are represented in the same colors as in Fig.\ \ref{figure-soluble}(a). The lactose disaccharides of the carbohydrate tips, which are located between the Le$^{\rm X}$ trisaccharide and the linker group of the glycolipid, are represented in light blue and light yellow, respectively. 
}
\label{figure-trans}
\end{figure*}

The interaction events of the two Le$^{\rm X}$ molecules or the two Lac 2 molecules can be characterized by their lifetime and by the maximum number of contacts of the events. These interaction events are obtained from our simulation trajectories as consecutive stretches of frames at intervals of $0.1$ ns with nonzero contacts of the two molecules. Fig.\ \ref{figure-soluble}(d) shows that the average lifetime of the interaction events increases with the maximum number of contacts observed during the event. With average lifetimes in the nanoseconds range, the interactions of the two Le$^{\rm X}$ or the two Lac 2 molecules are rather short-lived. Nonetheless, the radial distribution functions in Fig.\ \ref{figure-soluble}(e) indicate that the interactions are attractive.  The maxima of the radial distribution functions at center-of-mass distances of about $0.8$ nm for Le$^{\rm X}$ and $0.6$ nm for Lac 2 are significantly larger than the value 1 for a non-interacting ideal solution. 

Quantifying the attractive interactions of the two Le$^{\rm X}$ or two Lac 2 molecules requires distinguishing bound and unbound states. This distinction is somewhat arbitrary because of the fuzzy interactions of the carbohydrates. The probability distributions of carboyhydrate-carbohydrate contact numbers in Fig.\ \ref{figure-soluble}(c) are monotonously decreasing and, thus, not 
bimodal as required for a clear distinction of two states. Table 1 presents association constants of two Le$^{\rm X}$ or two Lac 2 molecules calculated for different cutoffs $n_c$ of the maximum number of contacts of interaction events. In these calculations, only interaction events with a maximum number of contacts larger or equal to the cutoff $n_c$ are taken to be binding events. The probability $P_b$ that the two Le$^{\rm X}$ or two Lac 2 molecules are bound has been determined from the total duration of the binding events, and the association constants from $K_a = V P_b/P_u$ where $P_u = 1 - P_b$ is the probability that the molecules are unbound, and $V$ is the volume of the simulation box. The $K_a$ values in Table 1 slightly decrease with increasing contact cutoff $n_c$ for binding events. For Le$^{\rm X}$, a $K_a$  value of 10 M$^{-1}$ has been obtained from weak affinity chromatography experiments \cite{Fuente04}, which is of the same order of magnitude as the values derived from our simulations. 

\subsection*{Interactions of lipid-anchored carbohydrates}

To investigate the interactions of two lipid-anchored Le$^{\rm X}$ or two lipid-anchored Lac 2 molecules, we have performed simulations of Le$^{\rm X}$ and Lac 2 glycolipids embedded in POPC lipid membranes. Our Le$^{\rm X}$ and Lac 2 glycolipids  have the same lipid tails as POPC, and carbohydrate tips that are connected to these lipid tails by a glycerol linker group (see Fig.\ \ref{figure-structure}). The carbohydrate tip of the Le$^{\rm X}$  glycolipid consists of the Le$^{\rm X}$ trisaccharide and an additional lactose disaccharide as spacer between Le$^{\rm X}$ and the glycerol linker. The Lac 2 glycolipid has the linear Lac 2 tetrasaccharide as carbohydrate tip. The force field of our simulations combines the GLYCAM06$^{\rm TIP5P}_{\rm OSMOr14}$ carbohydrate force field \cite{Sauter16,Kirschner08} for the TIP5P water model with the AMBER Lipid14 force field \cite{Dickson14} for lipid membranes. Because simulations of AMBER Lipid14 POPC membranes in TIP5P water lead to an unreasonably small area per lipid, we have rescaled the Lennard-Jones interactions between the TIP5P water molecules and the lipid headgroup atoms to obtain the same area per lipid as in standard AMBER Lipid14 simulations with the TIP3P water model (see Methods). 

\begin{figure}[t]
\centering
\includegraphics[width=0.95\linewidth]{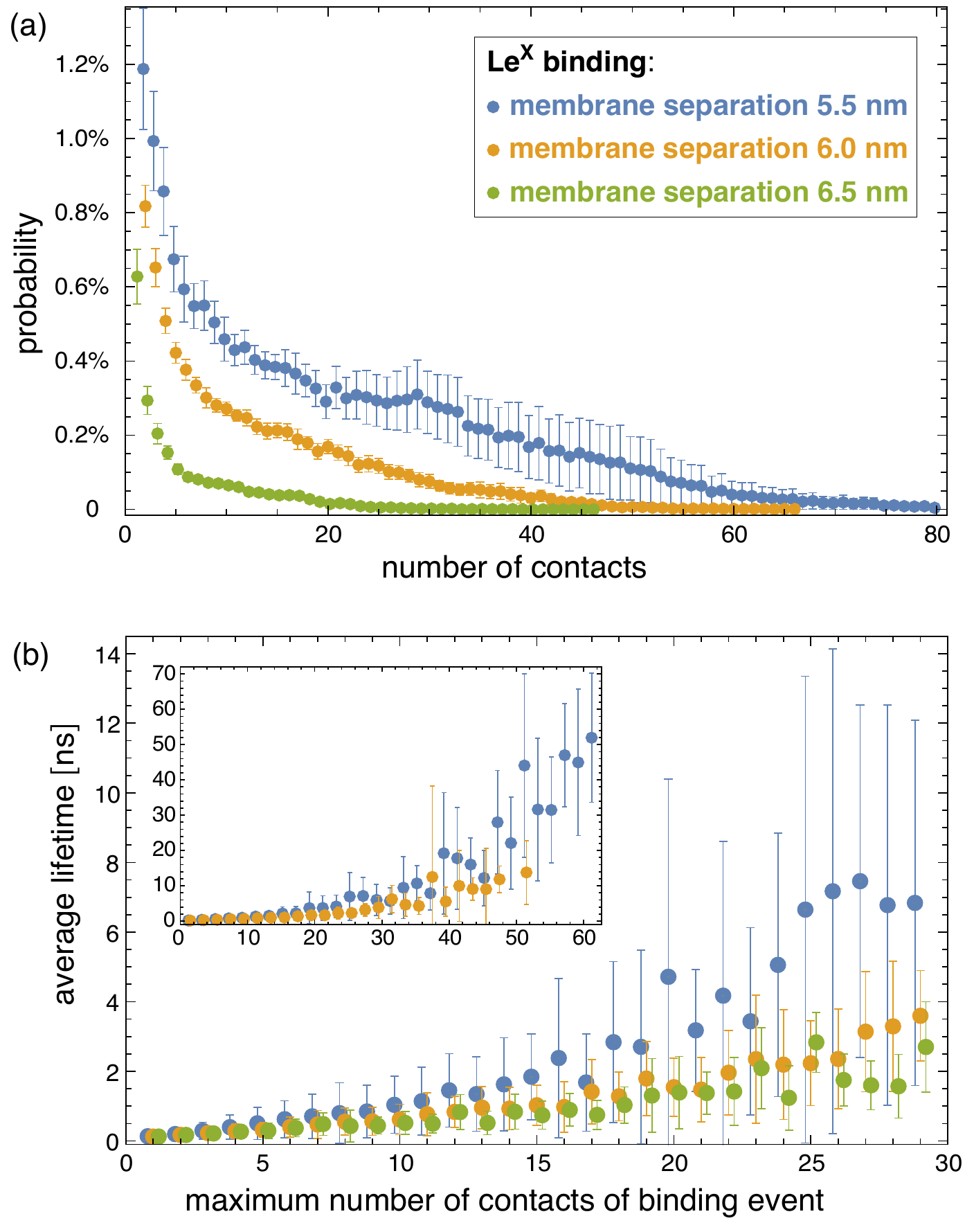}
\caption{(a) Probability distributions of the number of contacts between two Le$^{\rm X}$ glycolipids at different membrane separations. (b) Average lifetime of interaction events  at different membrane separations as a function of the maximum contact number of the events. 
}
\label{figure-trans-II}
\end{figure}

We quantify the interactions of two Le$^{\rm X}$ or two Lac 2 glycolipids at apposing membrane surfaces in a system that consists of a single lipid bilayer with one glycolipid anchored in each monolayer (see Fig.\ \ref{figure-trans}). In this system, the two glycolipids in the different monolayers interact due to the periodic boundary conditions of the simulation box, and the separation of the membrane monolayers can be adjusted by varying the number of water molecules in the simulation box. The values for the membrane separation $l$ given in Fig.~\ref{figure-trans} correspond to the separation from membrane midplane to membrane midplane and, thus, to the height of the simulation box. At each membrane separation, we have generated 10 trajectories with a length of  $3$ $\mu$s for the Le$^{\rm X}$ system and a length of $1$ $\mu$s for the Lac 2 system at the temperature $30\degree$C. The total simulation times at each membrane separation thus are $30$ $\mu$s and $10$ $\mu$s for the Le$^{\rm X}$ and Lac 2 systems, respectively. The membranes contain in each monolayer 35 lipids besides the single glycolipid and have an area $A$ of $23.3$ nm$^2$. { The height of the simulation box $l$ increases with the number of water molecules $n_w$ as $l \simeq 3.8\, {\rm nm}  +  0.013\, n_w \,  {\rm nm}$. The thickness of the water layer in the simulations thus is about $l- 3.8$ nm.}

The interactions of the glycolipids strongly depend on the membrane separation. For the membrane separations $l=5.5$, $6.0$, $6.5$, and $7.0$ nm, 50 randomly selected complexes of the Le$^{\rm X}$ glycolipid tips with at least 10 contacts of non-hydrogen atoms are displayed at the bottom of Fig.\ \ref{figure-trans}. The carbohydrate tip of the lower Le$^{\rm X}$ glycolipid is aligned in the 50 complexes and represented in blue colors, while the carbohydrate tip of the upper glycolipid is represented in red/yellow colors. The clouds of red/yellow carbohydrates illustrate that the interactions of lipid-anchored  Le$^{\rm X}$ are fuzzy, similar to soluble Le$^{\rm X}$ and Lac 2 (see Fig.\ \ref{figure-soluble}). The overlap of the cloud of the upper, red/yellow carbohydrates with the lower, blue carbohydrate decreases with increasing membrane separation. At the membrane separation $5.5$ nm, the Le$^{\rm X}$ glycolipids interact {\em via} their entire carbohydrate tips. At the separation $6.0$ nm, the interactions are limited to the Le$^{\rm X}$ trisaccharide of the glycolipid tip, and at the membrane separations $6.5$ nm and $7.0$ nm, the interactions are further restricted to the galactose and fucose monosaccharides at the branched end of the  Le$^{\rm X}$ glycolipid. The decrease of interactions with increasing separation is also reflected in the probability distributions of contact numbers shown in Fig.\ \ref{figure-trans-II}(a) and in the average lifetime of the interaction events for different maximum numbers of contacts in Fig.\ \ref{figure-trans-II}(b). At the smallest membrane separation $5.5$ nm, complexes of Le$^{\rm X}$ glycolipids can exhibit up to 60 and more contacts of non-hydrogen atoms (see Fig.\ \ref{figure-trans-II}(a)), and average lifetimes up to 50 ns for interaction events with a maximum number of 60 contacts (see inset of  Fig.\ \ref{figure-trans-II}(b)), which are about one order of magnitude larger than the average lifetimes for interaction events of soluble Le$^{\rm X}$ molecules with the same maximum number of contacts. At the membrane separations  $6.0$ and $6.5$ nm, the overall contact numbers and lifetimes of interaction events are significantly smaller. 

\begin{figure}[t]
\centering
\includegraphics[width=0.9\linewidth]{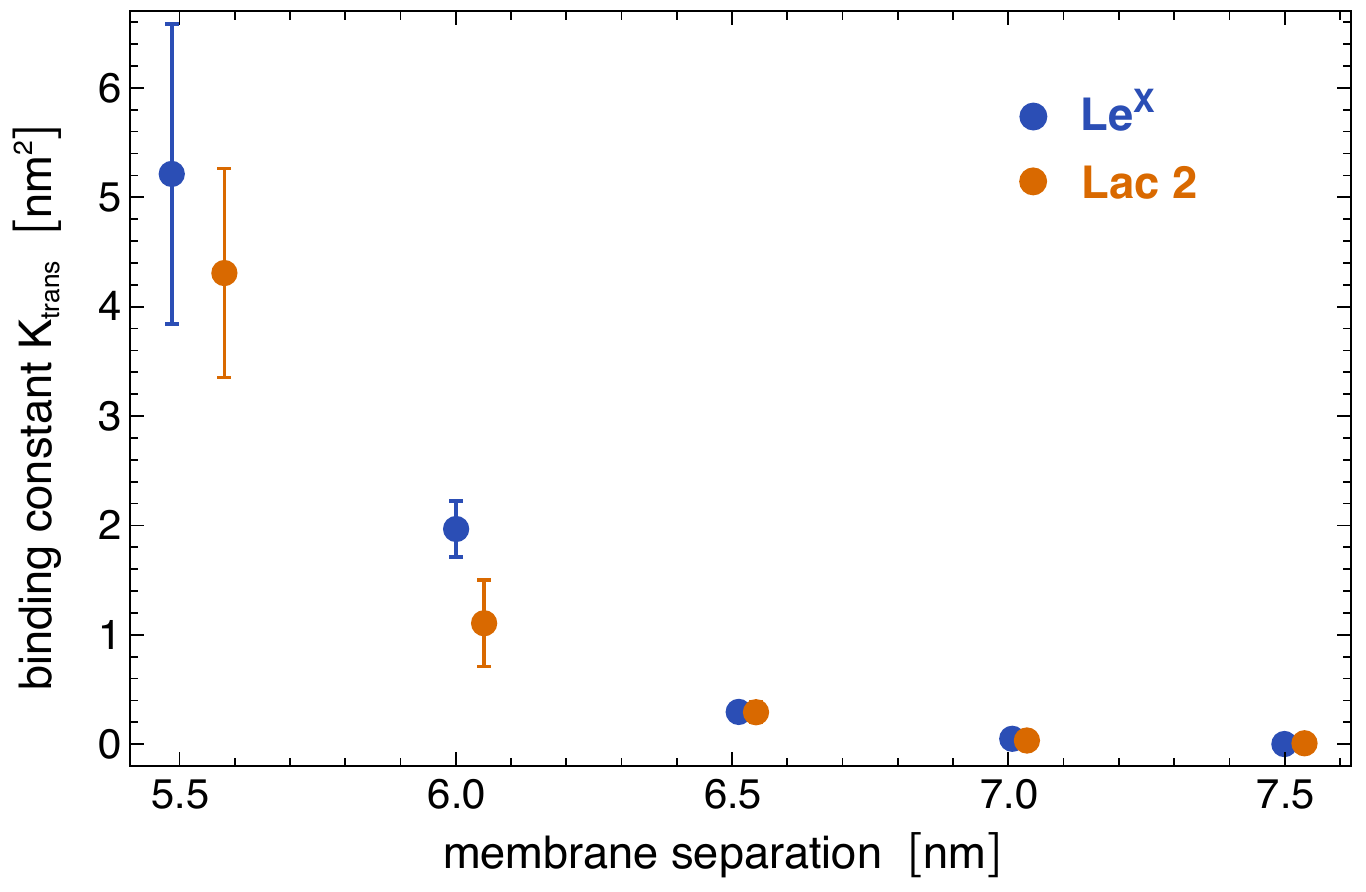}
\caption{Binding constant $K_{\rm trans}$ of two Le$^{\rm X}$ and two Lac 2 glycolipids {\it versus} membrane separation, calculated for binding events with a maximum number of at least $n_c= 5$ contacts of non-hydrogen atoms.  
}
\label{figure-K2D}
\end{figure}

Analogous to soluble carbohydrates, the binding association constants $K_{\rm trans}  = A P_b/(1-P_b)$ of the glycolipids in the different membrane monolayers can be determined from the probability $P_b$ that the two Le$^{\rm X}$ or two Lac 2 glycolipids are bound. The binding constants shown in Fig.\ \ref{figure-K2D} are calculated for binding events with a maximum number of at least $n_c= 5$ contacts of non-hydrogen atoms.  For the larger binding cutoff $n_c = 10$, the  $K_{\rm trans}$ values of the two Le$^{\rm X}$ glycolipids are about 10\% smaller than the values in Fig.\ \ref{figure-K2D} at the membrane separations $5.5$ and $6.0$ nm, and the values of the Lac 2 glycolipids are about 15\% smaller at these separations. The $K_{\rm trans}$ values decrease with increasing membrane separation. For membrane separations larger than about $7.5$ nm, the glycolipids cannot form contacts.  

The binding constant $K_{\rm trans}$ can be related to membrane adhesion energies, which have been measured for membrane vesicles that contain $10$ mol\% of Le$^{\rm X}$ glycolipids \cite{Pincet01,Gourier05}. For two apposing, large membrane surfaces of area $A$ that contain a total number of $N_t$ glycolipids, the free energy difference for forming the $n$th bond of the glycolipids is (see Methods)
\begin{equation}
\Delta G_n = - k_B T \ln[K_{\rm trans} (N_t - n + 1)^2/ n A]
\label{eq_Gn}
\end{equation}
The free energy differences $\Delta G_n$ are negative and, thus, favourable, from bond 1 until the equilibrium number $n_{\rm eq}$ of bonds. For bond numbers $n >n_{\rm eq}$, the free energy difference $\Delta G_n$ is positive and, thus, unfavorable for binding. The adhesion free energy $g_{\rm ad}$ per area now can be calculated by summing up the free energy differences $\Delta G_n$ from bond 1 to bond $n_{\rm eq}$:
\begin{equation}
g_{\rm ad} = \sum_{1}^{n_{\rm eq}} \Delta G_n / A
\label{eq_gad}
\end{equation}
For an area per lipid of $0.65$ nm$^2$ measured in our simulations, the area of a membrane surface that contains $N_t$ glycolipids at a concentration $10$ mol\% is $A \simeq 6.57 \, N_t \, {\rm nm}^2$.  From Eqs.\ (\ref{eq_Gn}) and (\ref{eq_gad}) and the values of $K_{\rm trans}$ for the Le$^{\rm X}$ glycolipids in Fig.\ \ref{figure-K2D}, we obtain the adhesion free energies $g_{\rm ad} = 320 \pm 60$, $150\pm 20$,  $28\pm 5$, and $5\pm 2$ $\mu$J/m$^2$ at the membrane separations $l=5.5$, $6.0$, $6.5$, and $7.0$ nm respectively. For lipid vesicles that contain $10$ mol\% of Le$^{\rm X}$ glycolipids, an adhesion free energy per area of $27\pm 2$ $\mu$J/m$^2$ has been reported \cite{Gourier05}, which is comparable to the adhesion free energy obtained from our simulations with membrane separation  $6.5$ nm.

\subsection*{Forces on lipid-anchored carbohydrates in trans-direction}

\begin{figure}[t]
\centering
\includegraphics[width=0.9\linewidth]{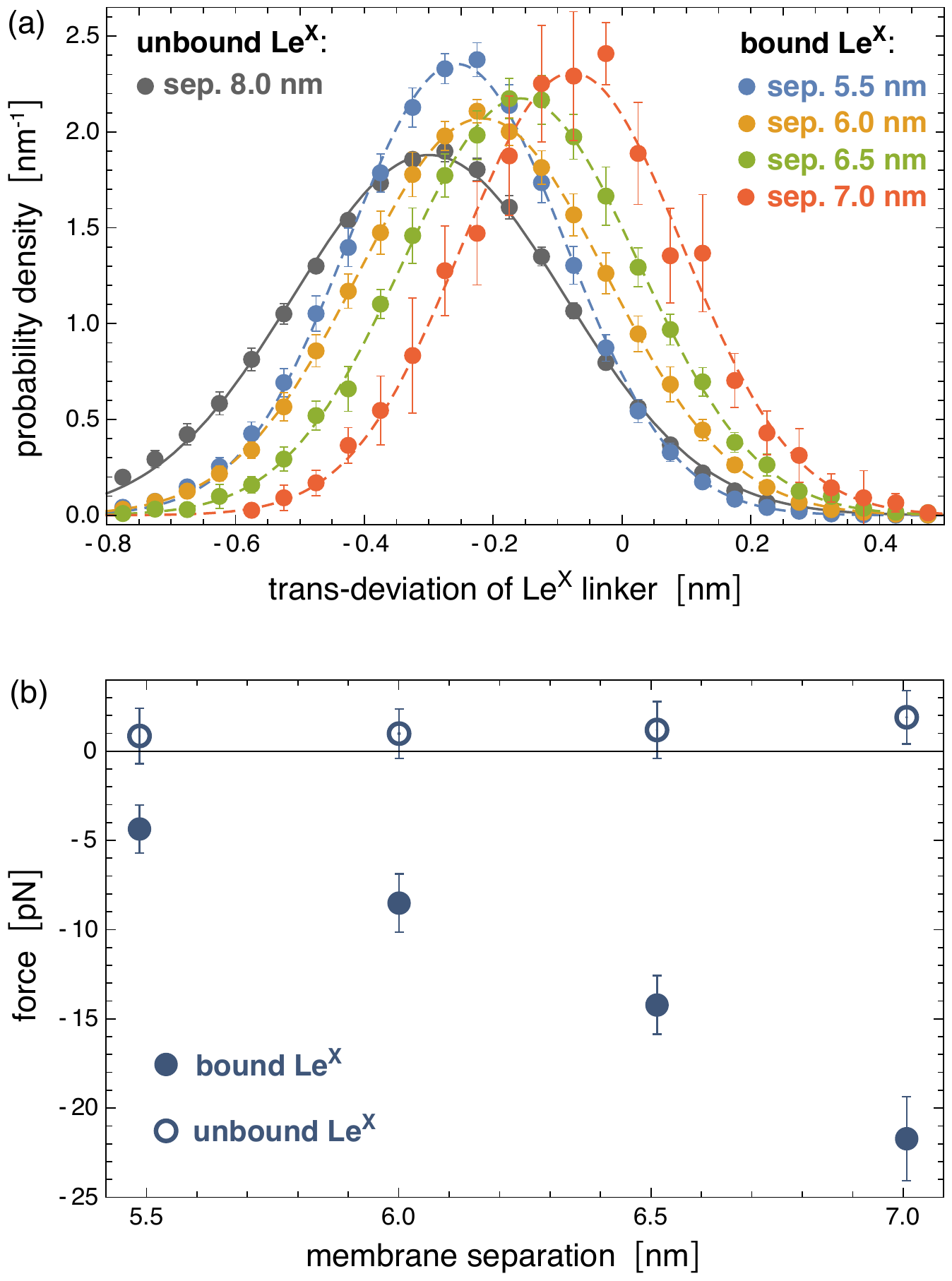}
\caption{(a) Distributions of trans-deviations of Le$^{\rm X}$ linker groups relative to the surrounding lipids. The trans-deviations are calculated as the difference between the center of mass of Le$^{\rm X}$ glycolipid linker group (see Fig.~\ref{figure-structure}) and the center of mass of all lipid head groups in the same monolayer as the glycolipid. These trans-deviations of Le$^{\rm X}$  in the direction perpendicular to the membrane plane are determined from the simulation trajectories of the system illustrated in Fig.~\ref{figure-trans}. (b) Forces on bound and unbound Le$^{\rm X}$ glycolipids at the different membrane separations. The trans-deviations and forces of bound glycolipids are obtained from the simulation frames of binding events with a maximum number of at least $n_c= 5$ contacts of non-hydrogen atoms. Deviations to force values obtained for the cutoff $n_c= 10$ are smaller than the error bars. Forces an unbound glycolipids are calculated from simulation frames with zero contacts between the glycolipids.}
\label{figure-forces}
\end{figure}

The binding of glycolipids in our simulations is associated with deviations of the glycolipids relative to the surrounding lipids. These deviations in the trans-direction perpendicular to the membrane surface result from forces on bound glycolipid complexes. Fig.\ \ref{figure-forces}(a) illustrates distributions of trans-deviations between the center of mass of the linker group  of a Le$^{\rm X}$ glycolipid (see Fig.\ \ref{figure-structure}) and the center of mass of all lipid head groups in the same monolayer  as the glycolipid. The trans-deviations $d$ are calculated from the simulation frames of our trajectories at intervals of $0.1$ ns. We obtain two values of $d$ per simulation frame for the two glycolipids relative to the monolayer in which they are embedded. An increase in $d$ indicates glycolipid motion away from the membrane midplane. With increasing membrane separation, the distributions for bound Le$^{\rm X}$ glycolipids deviate more and more from the distribution for unbound  Le$^{\rm X}$, which reflects increasing forces. The distribution of trans-deviations $d$ of unbound Le$^{\rm X}$ glycolipids shown in Fig.\ \ref{figure-forces}(a)  is calculated from our simulation trajectories at the membrane separation $8.0$ nm, at which Le$^{\rm X}$ bonds do not occur, and can be approximated by a Gaussian distribution $\exp[-V(d)/k_B T]$ with $V(d) = \frac{k}{2}(d - d_u)^2$. The trans-deviations $d$ of unbound  Le$^{\rm X}$ glycolipids thus can be described by a harmonic potential $V(d)$ with force constant $k$ and mean extension $d_{\rm u}$, which can be determined from the standard deviation $\sigma$ and mean $\bar{d}$ of the Gaussian as $k = k_BT/\sigma^2= 94 \pm 4 \, {\rm pN}/{\rm nm}$  and $d_{\rm u} = \bar{d} = -0.31 \pm 0.10$ nm.  The distributions of trans-deviations of bound Le$^{\rm X}$ glycolipids in Fig.\ \ref{figure-forces}(a) are calculated from our simulation trajectories at the membrane separations $5.5$, $6.0$, $6.5$, and $7.0$ nm, for binding events with a maximum number of at least $n_c= 5$ contacts of non-hydrogen atoms.
The average force $f = k(d_b -d_u)$ on bound Le$^{\rm X}$ glycolipids at the membrane separations $l=5.5$, $6.0$, $6.5$, and $7.0$ nm then can be calculated from the difference between the mean trans-deviations $d_b = -0.26\pm 0.01$, $-0.22\pm 0.01$, $-0.16\pm 0.01$, and $-0.08 \pm 0.02$ nm of the bound glycolipids at these membrane separations and the mean trans-deviation $d_u$ of the unbound glycolipids. The force $f$ on bound Le$^{\rm X}$ glycolipids  increases with increasing membrane separation up to a value of $21.7 \pm 2.4$ pN at the separation $7.0$ nm (see Fig.\ \ref{figure-forces}(b)). This maximal force value agrees with the unbinding force $20 \pm 4$ pN of two Le$^{\rm X}$ molecules obtained from atomic force microscopy experiments \cite{Tromas01}. For bound Lac 2 glycolipids, we obtain a maximal force of $14.7 \pm 3.5$ pN at the separation $7.0$ nm, which is about of the same magnitude as the maximal force sustained by the Le$^{\rm X}$ complexes.

\begin{figure}[t]
\centering
\includegraphics[width=0.9\linewidth]{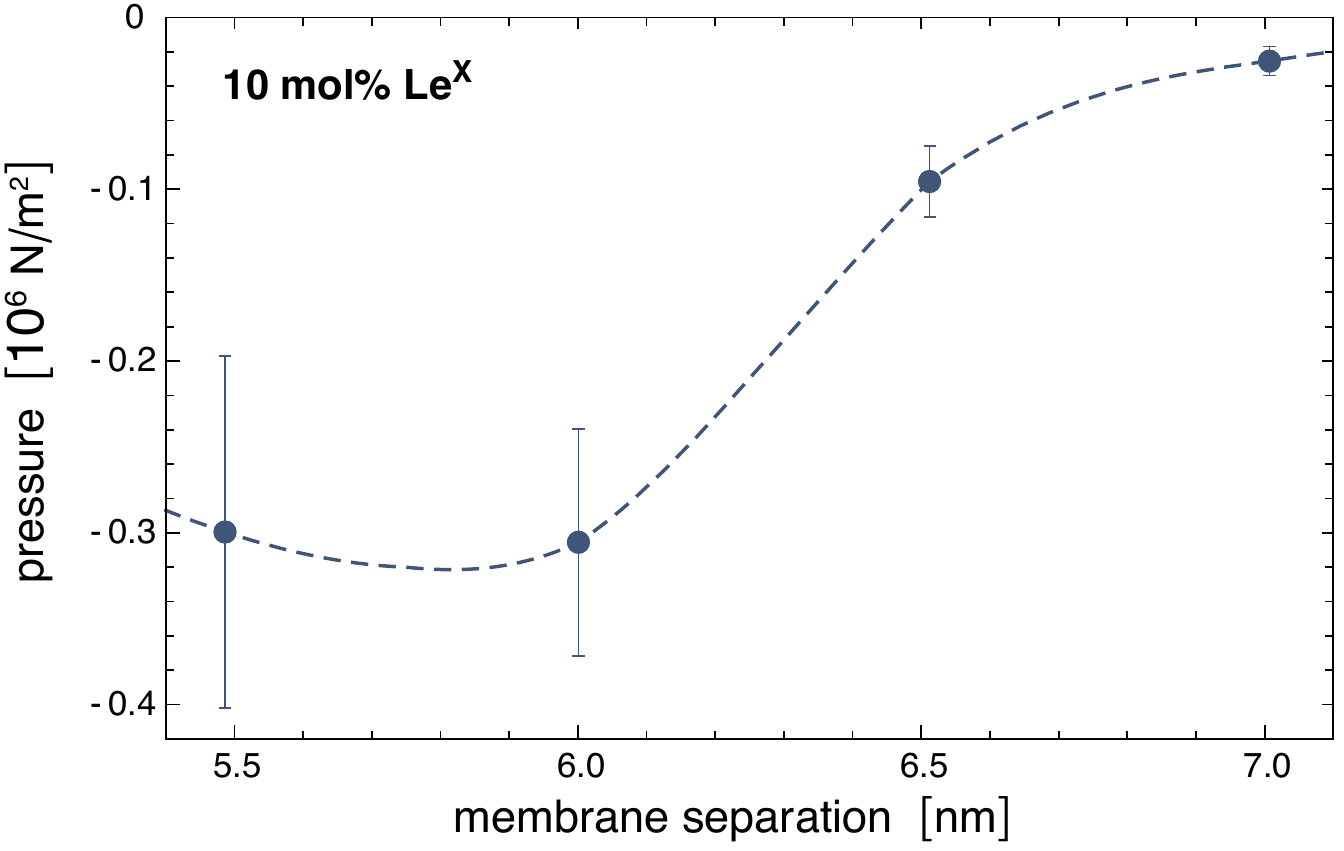}
\caption{Adhesion pressure $p$ of membranes with 10 mol\% of Le$^{\rm X}$ glycolipids obtained for the force values $f$ on bound  Le$^{\rm X}$ of Fig.\ \ref{figure-forces}(b). The dashed interpolation line is added as a guide for the eye and used to estimate adhesion energies  {\em via} integration (see text). In this integration, the pressure $p$ is taken to be zero at separations $l\ge 7.5$ nm.
}
\label{figure-pressure}
\end{figure}

The forces on bound  Le$^{\rm X}$ glycolipids lead to an adhesion pressure between the membranes. Fig.\ \ref{figure-pressure} illustrates the adhesion pressure $p$ of membranes that contain 10 mol\% of Le$^{\rm X}$ glycolipids as a function of the membrane separation. The adhesion pressure is estimated as $p = P_b f/A$ where $P_b$ is the probability that a Le$^{\rm X}$ glycolipid is bound at the concentration 10 mol\%,  $f$ is the average force on the bound glycolipid,  and $A \simeq 6.5$ nm$^2$ is the average membrane area of membrane patch with a single glycolipid at this concentration (see above). The negative pressure values for membrane separations $l$ of 7.0 nm and smaller, at which the glycolipids can bind, indicate membrane attraction. From integration of the pressure profile along the dashed interpolation line shown in Fig.\ \ref{figure-pressure}, we obtain adhesion energies  $g_{\rm ad} = \int_\infty^l p(l'){\rm d} l'  \simeq 140$ $\mu$J/m$^2$ for $l = 6.0$ nm and  $g_{\rm ad}  \simeq 30$  $\mu$J/m$^2$ for $l = 6.5$ nm. These adhesion energies per area agree with values $g_{\rm ad} = 150\pm 20$ $\mu$J/m$^2$ and  $28\pm 5$  $\mu$J/m$^2$ obtained directly from the binding constants $K_{\rm trans}$ at the membranes separations $l = 6.0$ and $6.5$ nm  (see above), which indicates that average forces $f$ on bound  Le$^{\rm X}$ glycolipids of Fig.\ \ref{figure-forces}(b) are consistent with the binding constants $K_{\rm trans}$ shown in Fig.\ \ref{figure-K2D}.

\section*{Discussion and Conclusions}

The membranes in our simulation systems are essentially planar because of the small size of the membranes, and because the glycolipid in one monolayer interacts with the glycolipid in the other monolayer across the periodic boundary of the simulation box.  In larger, experimental systems, in contrast, the membranes exhibit thermally excited shape fluctuations, which lead to a steric repulsion between adjacent membranes  \cite{Helfrich78}. During membrane adhesion, this steric repulsion needs to be overcome by attractive interactions \cite{Lipowsky86}. The average separation and thermal roughness of the adhering membranes is determined by the the interplay of the attractive interactions and the steric repulsion \cite{Steinkuhler19}. From neutron scattering experiments of DPPC membrane multilayers that contain 10 mol\% of Le$^{\rm X}$ glycolipids\cite{Schneck11}, an average membrane separation of $\bar{l} = 7.7 \pm 0.1$ nm and a relative membrane roughness of $\xi_\perp= 0.73\pm 0.03$ nm has been obtained.\footnote{The relative membrane roughness follows from Eq.\ (2) of Ref.~\citenum{Schneck11} as $\xi_\perp = \sqrt{g_1(0)}$ with parameter values given in Table 2. Here, $g_1(r)$ is the membrane displacement correlation function $g_1(r)$ of adjacent membranes in the multilayer.} Because of the periodicity of the membrane multilayers, the distribution of the local membrane separations $l$ between adjacent membranes can be approximated by the symmetric Gaussian distribution $P(l) \simeq \exp\left[-(l-\bar l)^2/2\xi_\perp^2\right]/(\sqrt{2\pi} \xi_\perp)$ with mean $\bar{l}$ and standard deviation $\xi_\perp$. The average membrane separation $\bar{l}$ obtained from neutron scattering is larger than the membrane separations at which the  Le$^{\rm X}$ glycolipids interact in our simulations. Trans-binding of the glycolipids therefore requires local membrane separations of the fluctuating membranes that are smaller than the average separation of the membranes. The average adhesion energy per area of adjacent membranes can be estimated as $\bar{g}_{\rm ad} = \int g_{\rm ad}(l) P(l) {\rm d} l$, where $g_{\rm ad}(l)$ is the adhesion energy as a function of the local membrane separation $l$. From the four values of $g_{\rm ad}(l)$ at the membrane separations $l=5.5$, $6.0$, $6.5$, and $7.0$ nm determined in the section "Interactions of lipid-anchored carbohydrates", we obtain the estimate $\bar{g}_{\rm ad}=7\pm 3$ $\mu$J/m$^2$ for the average separation $\bar{l}$ and relative membrane roughness $\xi_\perp$ of the neutron scattering experiments. This estimate of the average adhesion energy per area is comparable in magnitude to the adhesion free energy per area of $27\pm 2$ $\mu$J/m$^2$ reported for adhering membrane vesicles that contain 10 mol\% of Le$^{\rm X}$ glycolipids \cite{Gourier05}. The Le$^{\rm X}$ glycolipids embedded in the vesicles have the same carbohydrate tip as the Le$^{\rm X}$ glycolipids of the neutron scattering experiments and of our simulations. However, the carbohydrate tip of the vesicle system is connected to a ceramide, which contains a different linker between the carbohydrate tip and the lipid tails. Another difference is that the neutron scattering experiments have been performed at the temperature 60$\degree$C to ensure that the DPPC membranes in these experiments are fluid \cite{Schneck11}. The Le$^{\rm X}$ glycolipids of our simulations differ from those of the neutron scattering experiments only in the lipid tails. We have focused on POPC membranes and corresponding glycolipid tails to be able to run simulations of fluid membranes at the temperature 30$\degree$C, which is close to the calibration temperature of the force fields.  
In principle, membrane tension suppresses shape fluctuations of the membranes and can lead to stronger adhesion. However, the suppression of fluctuations occurs only on lateral length scales larger than the characteristic length $\sqrt{\kappa/\sigma}$\cite{Weikl18}, which adopts values between 100 and 400 nm for typical membrane tensions $\sigma$ of a few $\mu\text{N}/\text{m}$ \cite{Simson98,Popescu06,Betz09} and typical membrane bending rigidities $\kappa$ between $10$ and $40$ $k_B T$.\cite{Nagle13,Dimova14}. These values are significantly larger than the lateral correlation length $\xi_\parallel$ of membranes adhering {\em via} Le$^{\rm X}$ glycolipids, which is only a few nanometers for a relative membrane roughness $\xi_\perp$ of about $0.7$ nm \cite{Xu15}. On these small length scales, the membrane shape fluctuations are dominated by the bending energy of the membranes and the adhesion energies of  the glycolipids, and are not affected by membrane tension. 

The fuzzy interactions and comparable magnitude of the association constants of Le$^{\rm X}$ and Lac 2 obtained in our simulations indicates that the interactions of small, neutral carbohydrates  such as Le$^{\rm X}$ and Lac 2 are rather generic and not dependent on specific, structural aspects of the carbohydrates. The good agreement to experimental results for the association constant of soluble Le$^{\rm X}$ \cite{Fuente04}, adhesion energies of membranes with Le$^{\rm X}$ glycolipids \cite{Gourier05}, and maximally sustained forces of Le$^{\rm X}$ complexes \cite{Tromas01} suggests that our simulations provide a realistic, detailed picture of weak carbohydrate-carbohydrate interactions in solution as well as in membrane adhesion.
 The fuzzy binding reduces the loss of rotational and translational entropy of the molecules during binding \cite{Tompa08}, because binding can occur for a large variety of different relative orientations of the saccharides, in contrast to e.g.\ binding {\em via} specific hydrogen-bond patterns as suggested previously for Le$^{\rm X}$ based on simulations on short timescales up to 40 ns \cite{Gourmala07,Luo08}.
The fuzzy binding results from a subtle interplay between the rotational and translational entropy of the saccharides and the van der Waals, hydrogen bond, and hydrophobic interactions of the saccharides in the various binding conformations.

We have investigated the binding of Le$^{\rm X}$ in the absence of Ca$^{2+}$. Several groups have reported that Le$^{\rm X}$ binding depends on Ca$^{2+}$\cite{Geyer00,Fuente01,Hernaiz02,Gourier05,Nodet07,Kunze13,Witt16}, whereas other groups have observed no dependence on Ca$^{2+}$ \cite{Tromas01,Schneck11}. As pointed out by Kunze et al.\cite{Kunze13}, the Ca$^{2+}$ concentration used by most groups are of the order of 10 mM and, thus, greatly beyond physiological Ca$^{2+}$ concentrations. In vesicle adhesion experiments, Kunze et al.\cite{Kunze13} observed a rather small increase of the number of bound vesicles for a physiological Ca$^{2+}$ concentration of 0.9 mM, compared to experiments in the absence of Ca$^{2+}$. However, a strong increase of the number of bound vesicles in the experiments occurred for a Ca$^{2+}$ concentration of 10 mM. In atomic force microscopy experiments of Le$^{\rm X}$ unbinding \cite{Tromas01}, in contrast, the same unbinding force of about $20 \pm 4$ pN has been obtained both in the absence of Ca$^{2+}$ and for a Ca$^{2+}$ concentration of 10 mM. Overall, these experimental results suggest that the binding of Le$^{\rm X}$  is not strongly affected at least by physiological concentrations of Ca$^{2+}$.

\section*{Methods}

%
\subsection*{Simulations of soluble carbohydrates}

{\bf System setup} -- We have used the GLYCAM06$^{\rm TIP5P}_{\rm OSMOr14}$ carbohydrate force field \cite{Sauter16,Kirschner08} in our simulations of soluble pairs of Le$^{\rm X}$ and Lac 2 in water.
Initial structures of the Le$^{\rm X}$ trisaccharides and  Lac 2 tetrasacchardies were created with the Glycam Carbohydrate Builder program \cite{glycamweb} and solvated in truncated octahedral simulation boxes with $4287$ TIP5P water molecules for the Le$^{\rm X}$ pair and with $8504$ TIP5P water molecules for the Lac 2 pair. In the initial conformations, the two saccharides were placed in the simulation boxes such that they were not in contact. We have subsequently minimized the simulation systems in $5000$ minimization steps of steepest decent and additional $5000$ steps of the conjugent gradient algorithm. The systems were then heated from the temperature 0 K to 303 K at constant volume in $50\, 000$ integration time steps of $2$ fs with temperature control by a Langevin thermostat \cite{Salomon13}  with collision frequency $\gamma$~=~1.0~ps$^{-1}$. 

{\bf Production runs} -- After equilibration for $2$ ns at $303$ K, we have generated $50$ independent trajectories for the Le$^{\rm X}$ pair and $40$ trajectories for the Lac 2 pair with a $2$ fs integration step in AMBER 14 and 16 GPU  \cite{Salomon-Ferrer13,Le-Grand13} using the Monte-Carlo barostat \cite{Allen17} and a Langevin thermostat with collision frequency $\gamma$= 1.0~ps$^{-1}$ to keep the temperature at $303$ K and the pressure at $1$ bar. On these trajectories, the lengths of bonds that contain hydrogens were restrained with the SHAKE algorithm \cite{Miyamoto92,Ryckaert77}, non-bonded interactions were truncated at a cutoff value of 1 nm, and the Particle Mesh Ewald algorithm (PME) \cite{Essmann95,Darden93} was used to treat all electrostatic interactions. The $50$ simulation trajectories for Le$^{\rm X}$ pair have a length of $2.0$ $\mu$s, and the $40$ trajectories for the Lac 2 pair have a length of 1  $\mu$s or close to 1  $\mu$s. The total simulation times of these trajectories are $100$ $\mu$s for Le$^{\rm X}$ system and $39.5$ $\mu$s for the Lac 2 system. 

{\bf Analysis of trajectories} -- We have identified interactions events of the two Le$^{\rm X}$  or two Lac 2 molecules along the simulation trajectories as  consecutive stretches of simulation frames at intervals of $0.1$ ns with nonzero contacts of the molecules. These interaction events are separated by stretches of simulation frames with zero contacts and can be characterized by their lifetime and by the maximum number of contacts during the events. The contacts are defined as contacts between non-hydrogen atoms of the two molecules within a distance of less than $0.45$ nm. We consider interaction events with a maximum number of contacts that is larger or equal to a cutoff number $n_c$  as binding events. For the cutoff numbers $n_c = 5$, $10$, and $20$, we have obtained $7253$, $4820$, and $2331$ binding events of the two Le$^{\rm X}$ molecules on all trajectories, and $2369$, $1573$, and $823$ binding events of the two Lac 2 molecules. We have thus observed dozens of binding and unbinding events on each trajectory, with binding and unbinding times that are significantly smaller than the trajectory lengths (see also Fig.\ \ref{figure-soluble}(d)). To ensure independence from the initial, unbound conformations of the trajectories, we have discarded the first $100$ ns on all trajectories in our calculations of the binding probablity $P_b$ of the two molecules, which is defined as the fraction of simulation frames belonging  to binding events. We have calculated $P_b$ for each trajectory and have determined the overall value and error of $P_b$ as mean and error of the mean of the values for all trajectories. The association constants $K_a$ reported in Table~1 were calculated from these binding probabilities {\em via} the relation $K_a = V P_b/(1- P_b)$  where $V$ is the simulation box volume \cite{Hu13}. The errors of $K_a$ are calculated by error propagation from the errors of $P_b$. The errors of the probability distributions and radial distribution functions in Fig.\ \ref{figure-soluble}(c) and (e) are calculated as error of the mean of the corresponding quantities for the individual trajectories.

\begin{figure}[t]
\includegraphics[width=0.95\linewidth]{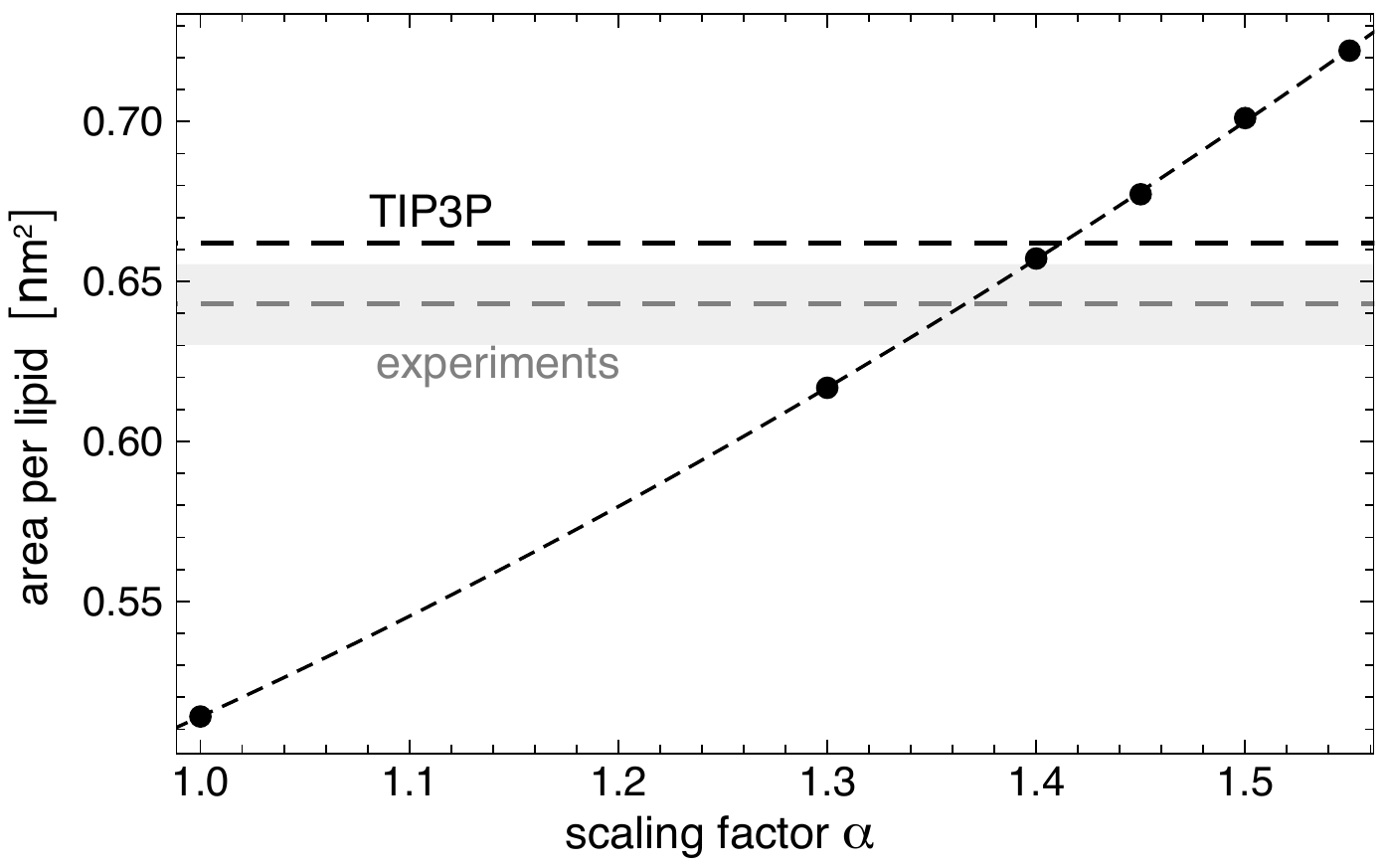}
\caption{Area per lipid for a Lipid14 POPC membrane in TIP5P water as a function of the scaling factor $\alpha$ for well-depth of the Lennard-Jones interactions between TIP5P water molecule and the lipid head group atoms. The dashed horizontal lines represent the area per lipid values of POPC membranes for Lipid14 in TIP3P water and from experiments \cite{Kucerka11}. The dashed line through the data points is a guide for the eye. Errors or the simulation data are smaller than the point sizes. The error of the experimental value is indicated by they shaded region. The temperature of the simulations and experiments is 30$\degree$C. 
}
\label{figure-area-per-lipid}
\end{figure}
\subsection*{Simulations of lipid-anchored saccharides}

{\bf System setup} -- We have  generated the initial structures of the POPC lipid membranes with the CHARMM-GUI program \cite{Jo08}. For our simulations with glycolipids, one lipid in each monolayer has been replaced by  a Le$^{\rm X}$ or Lac 2 glycolipid, which have the same lipids tails as POPC (see Fig.~\ref{figure-structure}). Following Ref.~\citen{Dickson14}, we have performed the initial minimization and equilibration steps of all membrane systems as follows: We have first performed a minimization of the water molecules for fixed lipids and glycolipids in $2500$ minimization steps of steepest descent and subsequent $2500$ steps of the conjugent gradient algorithm. The lipids and glycolipids have been fixed by harmonic constraints with a force constant of $500$ kcal mol$^{-1}$ \si{\angstrom}$^{-1}$ in this minimization. We have next removed the harmonic constraints, and have repeated the same minimization steps for the complete systems. The subsequent heating of the systems has been performed in three steps: (1) heating from $0$ K to $100$ K at constant volume with harmonic constraints on lipids and glycolipids with a force constant of $20$ kcal mol$^{-1}$ \si{\angstrom}$^{-1}$; (2) heating from $100$ K to $200$ K with a reduced force constant of $10$ kcal mol$^{-1}$ \si{\angstrom}$^{-1}$ of the harmonic constraints on lipids and glycolipids; and (3) heating from $200$ K to $303$ K at constant pressure and a membrane tension of zero with the same harmonic constraints as in the second step using a semi-isotropic pressure coupling and the  Berendsen barostat \cite{Berendsen84} with a pressure relaxation time of $3$ ps. Each heating step consist of $10\, 000$ MD integration steps of length $2$ fs with temperature control by a Langevin thermostat with a collision frequency of $5.0$ ps$^{-1}$.

\begin{figure}[t]
\includegraphics[width=0.95\linewidth]{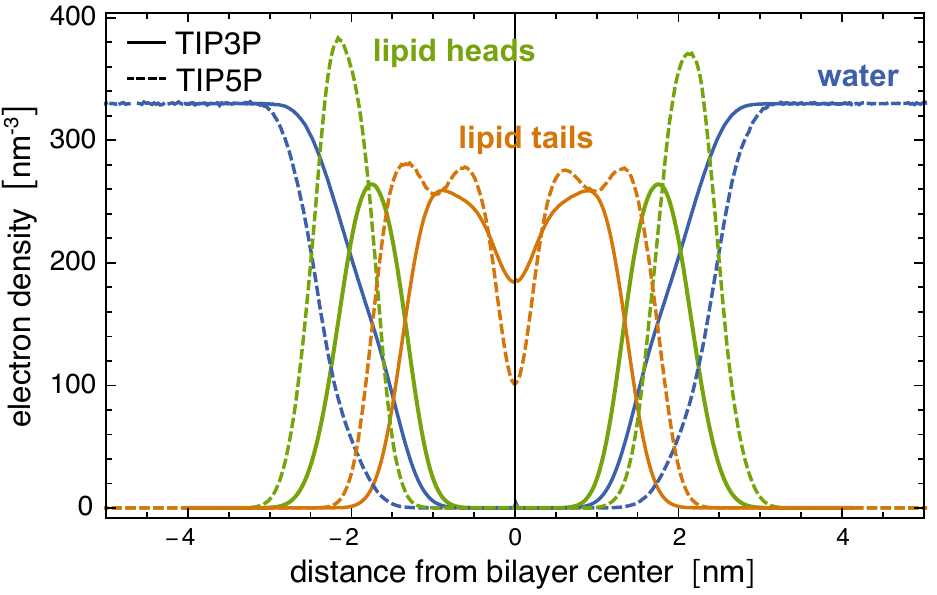}
\caption{Electron density profiles for a Lipid14 POPC membrane in TIP3P and TIP5P water at the temperature 303 K. The membrane is composed of 128 lipids.}
\label{figure-density-profiles}
\end{figure}

{\bf Rescaling of Lennard-Jones interactions between water and lipid headgroups} -- We have used the GLYCAM06$^{\rm TIP5P}_{\rm OSMOr14}$ carbohydrate force field \cite{Sauter16,Kirschner08} for the carbohydrates and the  AMBER Lipid14 force field \cite{Dickson14} for the lipids of our MD simulations of POPC membranes with glycolipids. Simulations of AMBER Lipid14 POPC membranes in TIP5P water lead to an unreasonably small area per lipid of $0.514 \pm 0.002$ (see Fig.\ \ref{figure-area-per-lipid}) and to density profiles that deviate significantly from profiles obtained from simulations in the standard TIP3P water model (see Fig.\ \ref{figure-density-profiles}), which has been used in the parametrization of the AMBER Lipid 14 force field \cite{Dickson14}. We have therefore rescaled the well depth of the Lennard-Jones interactions between the TIP5P water molecules and the Lipid14 lipid headgroup atoms by a scaling factor $\alpha$ in order to obtain the same area per lipid as in simulations of POPC membranes with TIP3P water. We chose to rescale the Lennard-Jones interactions between water and lipid headgroups because the density profiles of AMBER Lipid14 POPC membranes in TIP5P water show a smaller overlap between the water and lipid head group regions, compared to TIP3P water (see Fig.\ \ref{figure-density-profiles}).  This smaller overlap likely results from weaker Lennard-Jones interactions, and not from different atom sizes, because the Lennard-Jones radius $3.502~\AA$ of the TIP5P oxygen atom is in fact smaller than the radius $3.53~\AA$ of the TIP3P oxygen atom. Therefore, we have only rescaled the Lennard-Jones well-depth $\epsilon$ for the interaction between TIP5P water and the lipid head group atoms by a scaling factor $\alpha$.

Fig.\ \ref{figure-area-per-lipid}) illustrates simulation results for the area per lipid as a function of  the scaling factor $\alpha$.  The membranes in these simulations consists of 128 POPC lipids, and the number of water molecules is 6400. For each value of $\alpha$, we have generated $10$ independent trajectories of length $150$ ns with semi-isotropic pressure coupling at a membrane tension of zero and a temperature of 303 K using the same barostat and thermostat settings as in the last heating step of the system setup (see above). We have determined the area per lipid from the last 100 ns of these trajectories, with errors calculated as error of the mean of the values for the individual trajectories. The value $\alpha = 1.4$  leads to an area per lipid in TIP5P simulations that is close to the area per lipid both in TIP3P simulations and in experiments (see Fig.\ \ref{figure-area-per-lipid}). We have therefore used $\alpha = 1.4$ in our simulations of lipid-anchored saccharides. For this value of  $\alpha$, the density profile of AMBER Lipid14 POPC membranes in TIP5P water (not shown) is practically identical to density profile in TIP3P water, and the membrane thickness $d_m$ and lateral diffusion coefficient $D$ of the lipids are identical within errors or close to the values obtained in TIP3P simulations (see Table \ref{table-scaling}). We have determined the bilayer thickness as the distance between the electron density peaks of the lipid head groups, and the lateral diffusion constant from the relation $D  = MSD(t)/(4 t)$ where $MSD(t)$ is the mean-squared-displacement of a lipid molecule at time $t$. To obtain $MSD(t)$, we have first removed the center of mass motion of each leaflet to eliminate the `caterpillar effect' \cite{Anezo03} and have divided our trajectories into $20$ ns fragments. We have then calculated  $MSD(t)$ from the MSD profiles of single lipids by averaging over all lipids and all trajectory fragments. The diffusion coefficients in Table \ref{table-scaling} are calculated from linear fits in the time intervals from $t = 10$ ns to $20$ ns in which $MSD(t)$ approaches a constant slope.

\begin{table}[t]
\centering
\caption{Membrane thickness $d_m$ and lipid diffusion coefficient $D$ from simulations with TIP5P water for different values of the scaling factor $\alpha$, from simulations with TIP3P water, and from experiments on POPC lipid membranes}  
\begin{tabular}{l | cc}
$\alpha$  & $d_m$ $[{\rm nm}]$ & $D \; [\mu {\rm m}^2/{\rm s}]$ \\ \hline
1.2   & $3.82 \pm 0.03$     &  $3.27 \pm 0.10$   \\
1.3   &  $3.67 \pm 0.01$     & $4.15 \pm 0.08$    \\
1.4   & $3.51 \pm 0.01$      &  $5.57 \pm 0.16$    \\
1.45  & $3.43 \pm 0.01$     &  $5.85 \pm 0.15$    \\
1.5    & $3.36 \pm 0.01$       &  $6.77 \pm 0.11$    \\
1.55  &  $3.31 \pm 0.01$     & $7.31 \pm 0.19$    \\ \hline
TIP3P &  $3.54 \pm 0.01$      &  $5.45 \pm 0.19$    \\
exp.\  &  $3.68$ \cite{Kucerka05} &  $10.7$ \cite{Filippov03} 
\end{tabular}
\label{table-scaling}
\end{table}

{\bf Production runs} --  The membranes of our simulations with two Le$^{\rm X}$ or two Lac 2 glycolipids are composed of 35 POPC lipids and one glycolipid in each monolayer. By varying the number of water molecules in the simulation box, we have created several membrane systems that differ in simulation box height. In our simulations with Le$^{\rm X}$ glycolipids, we have obtained the average box heights $l = 5.49$,  $6.00$, $6.51$, $7.01$, $7.50$, and $8.00$ nm for the numbers $1264$, $1722$, $2107$, $2493$, $2878$, and $3263$ of TIP5P water molecules, respectively. In our simulations with Lac 2 glycolipids, we have obtained the average box heights $l=5.58$, $6.05$,  $6.54$, $7.03$, and $7.53$ nm for the numbers $1373$, $1753$, $2132$, $2512$, and $2891$ of water molecules. The height $l$ of the rectangular simulation box corresponds to the separation from membrane midplane to membrane midplane across the periodic boundary of the box in the direction perpendicular to the membrane. After equilibration for $100$ ns, we have produced $10$ independent trajectories for each system with the software AMBER 16 GPU \cite{Salomon-Ferrer13,Le-Grand13}. The trajectories have a length of  3 $\mu$s for the Le$^{\rm X}$ systems and a length of 1 $\mu$s for the Lac 2 systems. We have regulated the simulation temperature of $303$ K using a Langevin thermostat\cite{Salomon13} with a collision frequency of 5.0~ps$^{-1}$, and have employed a semi-isotropic pressure coupling with a pressure of 1 bar in all directions,  which corresponds to a membrane tension of zero.  We have used the Berendsen barostat \cite{Berendsen84} with relaxation time $\tau = 3$ ps for the pressure regulation because of the stability of the semi-isotropic pressure coupling in AMBER 16 GPU in combination with this barostat. For large systems as considered here, the weak-coupling scheme of the Berendsen barostat can be expected to lead to results that are essentially equivalent to other barostats \cite{Morishita00}. We have constrained the bond lengths for hydrogen atoms with the SHAKE algorithm \cite{Miyamoto92,Ryckaert77} and have used an integration timestep of $2$ fs in all simulations. A cutoff length of $1.0$~nm was used in calculating the non-bonded interactions with the Particle Mesh Ewald (PME) algorithm \cite{Essmann95,Darden93}. 

{\bf Analysis of trajectories} -- We have identified interactions events between the carbohydrate tips of the two Le$^{\rm X}$ or two Lac 2 glycolipids in the same way as described above for the soluble saccharides. For two Le$^{\rm X}$ glycolipids, we have obtained $1490$,  $1609$, $588$, $141$ binding events with a maximum contact number of at least $n_c=5$ on the trajectories at the membrane separations $5.5$, $6.0$, $6.5$, and $7.0$ nm, respectively.  For two Lac 2 glycolipids, we have obtained $609$, $413$, $183$, and $34$ such binding events on the trajectories at the corresponding membrane separations. 

To ensure independence from the initial conformation of the trajectories, we have discarded the first 10\% of each trajectory in our calculations of the binding probablity $P_b$ of the two molecules. In analogy to soluble carbohydrates, we have determined $P_b$ and its error as mean and error of the mean of the values for the 10 trajectories at a given membrane separation. The binding constant then follows  as 
 $K_{\rm trans} = A P_b/(1- P_b)$  where $A$ is the membrane ara \cite{Hu13}. We have calculated  the errors of the probability distributions in Figs.\ \ref{figure-trans-II}(a) and \ref{figure-forces}(a) and of the forces in  Fig.\ \ref{figure-forces}(b) as error of the mean of the corresponding quantities for the individual trajectories.

\subsection*{Calculation of adhesion free energies from trans-binding constants of membrane-anchored molecules}

The binding constant $K_{\rm trans}$ of molecules anchored to two apposing membrane surfaces $1$ and $2$ of area $A$ is related to the on- and off-rate constants of these molecules {\em via}
\begin{equation}
K_{\rm trans} = k_{\rm on}/k_{\rm off}
\end{equation}
 If the total numbers of the molecules at the two surfaces are $N_1$ and $N_2$, up to  $n\le {\rm min}(N_1, N_2)$ trans-bonds can be formed. The effective rate for going from a state with $n-1$ trans-bonds to a state with $n$ bonds is \cite{Hu13}
\begin{equation}
k_+ = k_{\rm on} (N_1 - n +1)(N_2 -n +1) /A
\end{equation}
and the effective rate for going back from $n$ bonds to $n-1$ is
\begin{equation}
k_- = n k_{\rm off}
\end{equation}
The condition of detailed balance implies 
\begin{equation}
P_{n-1} k_+ = k_- P_n
\end{equation}
where $P_n$ is the equilibrium probability of the state with $n$ trans-bonds. The free-energy difference $\Delta G_n$ between the states with $n$ and $n-1$ bonds is related to the equilibrium probabilities {\em via}
\begin{equation}
 \exp[-\Delta G_n/k_B T] = P_n/P_{n-1}
\end{equation} 
From these equations, we obtain
\begin{equation}
\Delta G_n = - k_B T \ln\left[\frac{K_{\rm trans}(N_1 -n +1)(N_2 - n +1)}{n A}\right]
\end{equation}
The adhesion free energy $g_{\rm ad}$ per area then can be calculated by summing up the free energy differences $\Delta G_n$ from bond 1 to bond $n_{\rm eq}$ where $n_{\rm eq}$ is the equilibrium number of bonds at which $\Delta G_n$ changes sign (see Eq.~\ref{eq_gad}).

\section*{Conflicts of interest}
There are no conflicts to declare.

\section*{Acknowledgements}
Financial support of the International Max Planck Research School (IMPRS) on Multiscale Bio-Systems and by the German Research Foundation (DFG) {\em via} Emmy Noether grant SCHN 1396/1 is gratefully acknowledged. We would like to thank Mark Santer for helpful discussions.



\balance



\begin{mcitethebibliography}{70}
\providecommand*{\natexlab}[1]{#1}
\providecommand*{\mciteSetBstSublistMode}[1]{}
\providecommand*{\mciteSetBstMaxWidthForm}[2]{}
\providecommand*{\mciteBstWouldAddEndPuncttrue}
  {\def\EndOfBibitem{\unskip.}}
\providecommand*{\mciteBstWouldAddEndPunctfalse}
  {\let\EndOfBibitem\relax}
\providecommand*{\mciteSetBstMidEndSepPunct}[3]{}
\providecommand*{\mciteSetBstSublistLabelBeginEnd}[3]{}
\providecommand*{\EndOfBibitem}{}
\mciteSetBstSublistMode{f}
\mciteSetBstMaxWidthForm{subitem}
{(\emph{\alph{mcitesubitemcount}})}
\mciteSetBstSublistLabelBeginEnd{\mcitemaxwidthsubitemform\space}
{\relax}{\relax}

\bibitem[Alberts \emph{et~al.}(2014)Alberts, Johnson, Lewis, Morgan, Raff,
  Roberts, Walter, Wilson, and Hunt]{Alberts14}
B.~Alberts, A.~Johnson, J.~Lewis, D.~Morgan, M.~Raff, K.~Roberts, P.~Walter,
  J.~Wilson and T.~Hunt, \emph{Molecular Biology of the Cell, 6th ed.}, Garland
  Science, New York, 2014\relax
\mciteBstWouldAddEndPuncttrue
\mciteSetBstMidEndSepPunct{\mcitedefaultmidpunct}
{\mcitedefaultendpunct}{\mcitedefaultseppunct}\relax
\EndOfBibitem
\bibitem[Varki(2007)]{Varki07}
A.~Varki, \emph{Nature}, 2007, \textbf{446}, 1023--1029\relax
\mciteBstWouldAddEndPuncttrue
\mciteSetBstMidEndSepPunct{\mcitedefaultmidpunct}
{\mcitedefaultendpunct}{\mcitedefaultseppunct}\relax
\EndOfBibitem
\bibitem[Dennis \emph{et~al.}(2009)Dennis, Nabi, and Demetriou]{Dennis09}
J.~W. Dennis, I.~R. Nabi and M.~Demetriou, \emph{Cell}, 2009, \textbf{139},
  1229--1241\relax
\mciteBstWouldAddEndPuncttrue
\mciteSetBstMidEndSepPunct{\mcitedefaultmidpunct}
{\mcitedefaultendpunct}{\mcitedefaultseppunct}\relax
\EndOfBibitem
\bibitem[Schnaar(2004)]{Schnaar04}
R.~L. Schnaar, \emph{Arch. Biochem. Biophys.}, 2004, \textbf{426},
  163--72\relax
\mciteBstWouldAddEndPuncttrue
\mciteSetBstMidEndSepPunct{\mcitedefaultmidpunct}
{\mcitedefaultendpunct}{\mcitedefaultseppunct}\relax
\EndOfBibitem
\bibitem[Liu and Rabinovich(2005)]{Liu05}
F.~T. Liu and G.~A. Rabinovich, \emph{Nat. Rev. Cancer}, 2005, \textbf{5},
  29--41\relax
\mciteBstWouldAddEndPuncttrue
\mciteSetBstMidEndSepPunct{\mcitedefaultmidpunct}
{\mcitedefaultendpunct}{\mcitedefaultseppunct}\relax
\EndOfBibitem
\bibitem[Arnaud \emph{et~al.}(2013)Arnaud, Audfray, and Imberty]{Arnaud13}
J.~Arnaud, A.~Audfray and A.~Imberty, \emph{Chem. Soc. Rev.}, 2013,
  \textbf{42}, 4798--4813\relax
\mciteBstWouldAddEndPuncttrue
\mciteSetBstMidEndSepPunct{\mcitedefaultmidpunct}
{\mcitedefaultendpunct}{\mcitedefaultseppunct}\relax
\EndOfBibitem
\bibitem[Varki(2017)]{Varki17}
A.~Varki, \emph{Glycobiology}, 2017, \textbf{27}, 3--49\relax
\mciteBstWouldAddEndPuncttrue
\mciteSetBstMidEndSepPunct{\mcitedefaultmidpunct}
{\mcitedefaultendpunct}{\mcitedefaultseppunct}\relax
\EndOfBibitem
\bibitem[Poole \emph{et~al.}(2018)Poole, Day, von Itzstein, Paton, and
  Jennings]{Poole18}
J.~Poole, C.~J. Day, M.~von Itzstein, J.~C. Paton and M.~P. Jennings,
  \emph{Nat. Rev. Microbiol.}, 2018, \textbf{16}, 440--452\relax
\mciteBstWouldAddEndPuncttrue
\mciteSetBstMidEndSepPunct{\mcitedefaultmidpunct}
{\mcitedefaultendpunct}{\mcitedefaultseppunct}\relax
\EndOfBibitem
\bibitem[Eggens \emph{et~al.}(1989)Eggens, Fenderson, Toyokuni, Dean, Stroud,
  and Hakomori]{Eggens89}
I.~Eggens, B.~Fenderson, T.~Toyokuni, B.~Dean, M.~Stroud and S.~Hakomori,
  \emph{J. Biol. Chem.}, 1989, \textbf{264}, 9476--9484\relax
\mciteBstWouldAddEndPuncttrue
\mciteSetBstMidEndSepPunct{\mcitedefaultmidpunct}
{\mcitedefaultendpunct}{\mcitedefaultseppunct}\relax
\EndOfBibitem
\bibitem[Kojima \emph{et~al.}(1994)Kojima, Fenderson, Stroud, Goldberg,
  Habermann, Toyokuni, and Hakomori]{Kojima94}
N.~Kojima, B.~A. Fenderson, M.~R. Stroud, R.~I. Goldberg, R.~Habermann,
  T.~Toyokuni and S.-I. Hakomori, \emph{Glycoconjugate Journal}, 1994,
  \textbf{11}, 238--248\relax
\mciteBstWouldAddEndPuncttrue
\mciteSetBstMidEndSepPunct{\mcitedefaultmidpunct}
{\mcitedefaultendpunct}{\mcitedefaultseppunct}\relax
\EndOfBibitem
\bibitem[Handa \emph{et~al.}(2007)Handa, Takatani-Nakase, Larue, Stemmler,
  Kemler, and Hakomori]{Handa07}
K.~Handa, T.~Takatani-Nakase, L.~Larue, M.~P. Stemmler, R.~Kemler and S.-I.
  Hakomori, \emph{Biochem. Biophys. Res. Commun.}, 2007, \textbf{358},
  247--252\relax
\mciteBstWouldAddEndPuncttrue
\mciteSetBstMidEndSepPunct{\mcitedefaultmidpunct}
{\mcitedefaultendpunct}{\mcitedefaultseppunct}\relax
\EndOfBibitem
\bibitem[Misevic \emph{et~al.}(1987)Misevic, Finne, and Burger]{Misevic87}
G.~N. Misevic, J.~Finne and M.~M. Burger, \emph{J. Biol. Chem.}, 1987,
  \textbf{262}, 5870--5877\relax
\mciteBstWouldAddEndPuncttrue
\mciteSetBstMidEndSepPunct{\mcitedefaultmidpunct}
{\mcitedefaultendpunct}{\mcitedefaultseppunct}\relax
\EndOfBibitem
\bibitem[de~la Fuente \emph{et~al.}(2001)de~la Fuente, Barrientos, Rojas, Rojo,
  Ca{\~n}ada, Fern{\'a}ndez, and Penad{\'e}s]{Fuente01}
J.~M. de~la Fuente, A.~G. Barrientos, T.~C. Rojas, J.~Rojo, J.~Ca{\~n}ada,
  A.~Fern{\'a}ndez and S.~Penad{\'e}s, \emph{Angew. Chem. Int. Ed.}, 2001,
  \textbf{40}, 2257--2261\relax
\mciteBstWouldAddEndPuncttrue
\mciteSetBstMidEndSepPunct{\mcitedefaultmidpunct}
{\mcitedefaultendpunct}{\mcitedefaultseppunct}\relax
\EndOfBibitem
\bibitem[Hernaiz \emph{et~al.}(2002)Hernaiz, de~la Fuente, Barrientos, and
  Penades]{Hernaiz02}
M.~J. Hernaiz, J.~M. de~la Fuente, A.~G. Barrientos and S.~Penades,
  \emph{Angew. Chem.-Int. Edit.}, 2002, \textbf{41}, 1554--1557\relax
\mciteBstWouldAddEndPuncttrue
\mciteSetBstMidEndSepPunct{\mcitedefaultmidpunct}
{\mcitedefaultendpunct}{\mcitedefaultseppunct}\relax
\EndOfBibitem
\bibitem[de~la Fuente \emph{et~al.}(2005)de~la Fuente, Eaton, Barrientos,
  Menendez, and Penades]{Fuente05}
J.~M. de~la Fuente, P.~Eaton, A.~G. Barrientos, M.~Menendez and S.~Penades,
  \emph{J. Am. Chem. Soc.}, 2005, \textbf{127}, 6192--6197\relax
\mciteBstWouldAddEndPuncttrue
\mciteSetBstMidEndSepPunct{\mcitedefaultmidpunct}
{\mcitedefaultendpunct}{\mcitedefaultseppunct}\relax
\EndOfBibitem
\bibitem[Tromas \emph{et~al.}(2001)Tromas, Rojo, de~la Fuente, Barrientos,
  Garcia, and Penades]{Tromas01}
C.~Tromas, J.~Rojo, J.~M. de~la Fuente, A.~G. Barrientos, R.~Garcia and
  S.~Penades, \emph{Angew. Chem. Int. Edit.}, 2001, \textbf{40},
  3052--3055\relax
\mciteBstWouldAddEndPuncttrue
\mciteSetBstMidEndSepPunct{\mcitedefaultmidpunct}
{\mcitedefaultendpunct}{\mcitedefaultseppunct}\relax
\EndOfBibitem
\bibitem[Bucior \emph{et~al.}(2004)Bucior, Scheuring, Engel, and
  Burger]{Bucior04}
I.~Bucior, S.~Scheuring, A.~Engel and M.~M. Burger, \emph{J. Cell Biol.}, 2004,
  \textbf{165}, 529--537\relax
\mciteBstWouldAddEndPuncttrue
\mciteSetBstMidEndSepPunct{\mcitedefaultmidpunct}
{\mcitedefaultendpunct}{\mcitedefaultseppunct}\relax
\EndOfBibitem
\bibitem[Lorenz \emph{et~al.}(2012)Lorenz, {\'A}lvarez~de Cienfuegos, Oelkers,
  Kriemen, Brand, Stephan, Sunnick, Y{\"u}ksel, Kalsani, Kumar, Werz, and
  Janshoff]{Lorenz12}
B.~Lorenz, L.~{\'A}lvarez~de Cienfuegos, M.~Oelkers, E.~Kriemen, C.~Brand,
  M.~Stephan, E.~Sunnick, D.~Y{\"u}ksel, V.~Kalsani, K.~Kumar, D.~B. Werz and
  A.~Janshoff, \emph{J. Am. Chem. Soc.}, 2012, \textbf{134}, 3326--9\relax
\mciteBstWouldAddEndPuncttrue
\mciteSetBstMidEndSepPunct{\mcitedefaultmidpunct}
{\mcitedefaultendpunct}{\mcitedefaultseppunct}\relax
\EndOfBibitem
\bibitem[Witt \emph{et~al.}(2016)Witt, Savic, Oelkers, Awan, Werz, Geil, and
  Janshoff]{Witt16}
H.~Witt, F.~Savic, M.~Oelkers, S.~I. Awan, D.~B. Werz, B.~Geil and A.~Janshoff,
  \emph{Biophys. J.}, 2016, \textbf{110}, 1582--1592\relax
\mciteBstWouldAddEndPuncttrue
\mciteSetBstMidEndSepPunct{\mcitedefaultmidpunct}
{\mcitedefaultendpunct}{\mcitedefaultseppunct}\relax
\EndOfBibitem
\bibitem[Gourier \emph{et~al.}(2005)Gourier, Pincet, Perez, Zhang, Zhu, Mallet,
  and Sinay]{Gourier05}
C.~Gourier, F.~Pincet, E.~Perez, Y.~M. Zhang, Z.~Y. Zhu, J.~M. Mallet and
  P.~Sinay, \emph{Angew. Chem.-Int. Edit.}, 2005, \textbf{44}, 1683--1687\relax
\mciteBstWouldAddEndPuncttrue
\mciteSetBstMidEndSepPunct{\mcitedefaultmidpunct}
{\mcitedefaultendpunct}{\mcitedefaultseppunct}\relax
\EndOfBibitem
\bibitem[Kunze \emph{et~al.}(2013)Kunze, Bally, H{\"o}{\"o}k, and
  Larson]{Kunze13}
A.~Kunze, M.~Bally, F.~H{\"o}{\"o}k and G.~Larson, \emph{Sci. Rep.}, 2013,
  \textbf{3}, 1452\relax
\mciteBstWouldAddEndPuncttrue
\mciteSetBstMidEndSepPunct{\mcitedefaultmidpunct}
{\mcitedefaultendpunct}{\mcitedefaultseppunct}\relax
\EndOfBibitem
\bibitem[Yu \emph{et~al.}(1998)Yu, Calvert, and Leckband]{Yu98}
Z.~W. Yu, T.~L. Calvert and D.~Leckband, \emph{Biochemistry}, 1998,
  \textbf{37}, 1540--1550\relax
\mciteBstWouldAddEndPuncttrue
\mciteSetBstMidEndSepPunct{\mcitedefaultmidpunct}
{\mcitedefaultendpunct}{\mcitedefaultseppunct}\relax
\EndOfBibitem
\bibitem[Schneck \emph{et~al.}(2011)Schneck, Deme, Gege, and Tanaka]{Schneck11}
E.~Schneck, B.~Deme, C.~Gege and M.~Tanaka, \emph{Biophys. J.}, 2011,
  \textbf{100}, 2151--2159\relax
\mciteBstWouldAddEndPuncttrue
\mciteSetBstMidEndSepPunct{\mcitedefaultmidpunct}
{\mcitedefaultendpunct}{\mcitedefaultseppunct}\relax
\EndOfBibitem
\bibitem[Latza \emph{et~al.}(2020)Latza, Deme, and Schneck]{Latza20}
V.~M. Latza, B.~Deme and E.~Schneck, \emph{Biophys. J.}, 2020, \textbf{118},
  1602--1611\relax
\mciteBstWouldAddEndPuncttrue
\mciteSetBstMidEndSepPunct{\mcitedefaultmidpunct}
{\mcitedefaultendpunct}{\mcitedefaultseppunct}\relax
\EndOfBibitem
\bibitem[Day \emph{et~al.}(2015)Day, Tran, Semchenko, Tram, Hartley-Tassell,
  Ng, King, Ulanovsky, McAtamney, Apicella, Tiralongo, Morona, Korolik, and
  Jennings]{Day15}
C.~J. Day, E.~N. Tran, E.~A. Semchenko, G.~Tram, L.~E. Hartley-Tassell,
  P.~S.~K. Ng, R.~M. King, R.~Ulanovsky, S.~McAtamney, M.~A. Apicella,
  J.~Tiralongo, R.~Morona, V.~Korolik and M.~P. Jennings, \emph{Proc. Natl.
  Acad. Sci. USA}, 2015, \textbf{112}, E7266--E7275\relax
\mciteBstWouldAddEndPuncttrue
\mciteSetBstMidEndSepPunct{\mcitedefaultmidpunct}
{\mcitedefaultendpunct}{\mcitedefaultseppunct}\relax
\EndOfBibitem
\bibitem[Yu \emph{et~al.}(2019)Yu, Tyrikos-Ergas, Zhu, Fittolani, Bordoni,
  Singhal, Fair, Grafmueller, Seeberger, and Delbianco]{Yu19}
Y.~Yu, T.~Tyrikos-Ergas, Y.~Zhu, G.~Fittolani, V.~Bordoni, A.~Singhal, R.~J.
  Fair, A.~Grafmueller, P.~H. Seeberger and M.~Delbianco, \emph{Angew.
  Chem.-Int. Edit.}, 2019, \textbf{58}, 13127--13132\relax
\mciteBstWouldAddEndPuncttrue
\mciteSetBstMidEndSepPunct{\mcitedefaultmidpunct}
{\mcitedefaultendpunct}{\mcitedefaultseppunct}\relax
\EndOfBibitem
\bibitem[Pincet \emph{et~al.}(2001)Pincet, Le~Bouar, Zhang, Esnault, Mallet,
  Perez, and Sinay]{Pincet01}
F.~Pincet, T.~Le~Bouar, Y.~M. Zhang, J.~Esnault, J.~M. Mallet, E.~Perez and
  P.~Sinay, \emph{Biophys. J.}, 2001, \textbf{80}, 1354--1358\relax
\mciteBstWouldAddEndPuncttrue
\mciteSetBstMidEndSepPunct{\mcitedefaultmidpunct}
{\mcitedefaultendpunct}{\mcitedefaultseppunct}\relax
\EndOfBibitem
\bibitem[Patel \emph{et~al.}(2007)Patel, Harding, Ebringerova, Deszczynski,
  Hromadkova, Togola, Paulsen, Morris, and Rowe]{Patel07}
T.~R. Patel, S.~E. Harding, A.~Ebringerova, M.~Deszczynski, Z.~Hromadkova,
  A.~Togola, B.~S. Paulsen, G.~A. Morris and A.~J. Rowe, \emph{Biophys. J.},
  2007, \textbf{93}, 741--749\relax
\mciteBstWouldAddEndPuncttrue
\mciteSetBstMidEndSepPunct{\mcitedefaultmidpunct}
{\mcitedefaultendpunct}{\mcitedefaultseppunct}\relax
\EndOfBibitem
\bibitem[Schneider \emph{et~al.}(2001)Schneider, Mathe, Tanaka, Gege, and
  Schmidt]{Schneider01}
M.~F. Schneider, G.~Mathe, M.~Tanaka, C.~Gege and R.~R. Schmidt, \emph{J. Phys.
  Chem. B}, 2001, \textbf{105}, 5178--5185\relax
\mciteBstWouldAddEndPuncttrue
\mciteSetBstMidEndSepPunct{\mcitedefaultmidpunct}
{\mcitedefaultendpunct}{\mcitedefaultseppunct}\relax
\EndOfBibitem
\bibitem[Sauter and Grafm\"uller(2016)]{Sauter16}
J.~Sauter and A.~Grafm\"uller, \emph{J. Chem. Theory Comput.}, 2016,
  \textbf{12}, 4375--4384\relax
\mciteBstWouldAddEndPuncttrue
\mciteSetBstMidEndSepPunct{\mcitedefaultmidpunct}
{\mcitedefaultendpunct}{\mcitedefaultseppunct}\relax
\EndOfBibitem
\bibitem[Lay \emph{et~al.}(2016)Lay, Miller, and Elcock]{Lay16}
W.~K. Lay, M.~S. Miller and A.~H. Elcock, \emph{J. Chem. Theory Comput.}, 2016,
  \textbf{12}, 1401--1407\relax
\mciteBstWouldAddEndPuncttrue
\mciteSetBstMidEndSepPunct{\mcitedefaultmidpunct}
{\mcitedefaultendpunct}{\mcitedefaultseppunct}\relax
\EndOfBibitem
\bibitem[Woods(2018)]{Woods18}
R.~J. Woods, \emph{Chem. Rev.}, 2018, \textbf{118}, 8005--8024\relax
\mciteBstWouldAddEndPuncttrue
\mciteSetBstMidEndSepPunct{\mcitedefaultmidpunct}
{\mcitedefaultendpunct}{\mcitedefaultseppunct}\relax
\EndOfBibitem
\bibitem[Gourmala \emph{et~al.}(2007)Gourmala, Luo, Barbault, Zhang, Ghalem,
  Maurel, and Fan]{Gourmala07}
C.~Gourmala, Y.~Luo, F.~Barbault, Y.~Zhang, S.~Ghalem, F.~Maurel and B.~Fan,
  \emph{J. Mol. Struct. Theochem}, 2007, \textbf{821}, 22 -- 29\relax
\mciteBstWouldAddEndPuncttrue
\mciteSetBstMidEndSepPunct{\mcitedefaultmidpunct}
{\mcitedefaultendpunct}{\mcitedefaultseppunct}\relax
\EndOfBibitem
\bibitem[Luo \emph{et~al.}(2008)Luo, Barbault, Gourmala, Zhang, Maurel, Hu, and
  Fan]{Luo08}
Y.~Luo, F.~Barbault, C.~Gourmala, Y.~Zhang, F.~Maurel, Y.~Hu and B.~T. Fan,
  \emph{J. Mol. Model.}, 2008, \textbf{14}, 901--910\relax
\mciteBstWouldAddEndPuncttrue
\mciteSetBstMidEndSepPunct{\mcitedefaultmidpunct}
{\mcitedefaultendpunct}{\mcitedefaultseppunct}\relax
\EndOfBibitem
\bibitem[Santos \emph{et~al.}(2009)Santos, Carvalho~de Souza, Ca{\~n}ada,
  Mart{\'\i}n-Santamar{\'\i}a, Kamerling, and Jim{\'e}nez-Barbero]{Santos09}
J.~I. Santos, A.~Carvalho~de Souza, F.~J. Ca{\~n}ada,
  S.~Mart{\'\i}n-Santamar{\'\i}a, J.~P. Kamerling and J.~Jim{\'e}nez-Barbero,
  \emph{Chembiochem}, 2009, \textbf{10}, 511--9\relax
\mciteBstWouldAddEndPuncttrue
\mciteSetBstMidEndSepPunct{\mcitedefaultmidpunct}
{\mcitedefaultendpunct}{\mcitedefaultseppunct}\relax
\EndOfBibitem
\bibitem[de~la Fuente and Penades(2004)]{Fuente04}
J.~M. de~la Fuente and S.~Penades, \emph{Glycoconjugate J.}, 2004, \textbf{21},
  149--163\relax
\mciteBstWouldAddEndPuncttrue
\mciteSetBstMidEndSepPunct{\mcitedefaultmidpunct}
{\mcitedefaultendpunct}{\mcitedefaultseppunct}\relax
\EndOfBibitem
\bibitem[Sauter and Grafm\"uller(2015)]{Sauter15}
J.~Sauter and A.~Grafm\"uller, \emph{J. Chem. Theory Comput.}, 2015,
  \textbf{11}, 1765--1774\relax
\mciteBstWouldAddEndPuncttrue
\mciteSetBstMidEndSepPunct{\mcitedefaultmidpunct}
{\mcitedefaultendpunct}{\mcitedefaultseppunct}\relax
\EndOfBibitem
\bibitem[Salomon-Ferrer \emph{et~al.}(2013)Salomon-Ferrer, Goetz, Poole,
  Le~Grand, and Walker]{Salomon-Ferrer13}
R.~Salomon-Ferrer, A.~W. Goetz, D.~Poole, S.~Le~Grand and R.~C. Walker,
  \emph{J. Chem. Theory Comput.}, 2013, \textbf{9}, 3878--3888\relax
\mciteBstWouldAddEndPuncttrue
\mciteSetBstMidEndSepPunct{\mcitedefaultmidpunct}
{\mcitedefaultendpunct}{\mcitedefaultseppunct}\relax
\EndOfBibitem
\bibitem[Le~Grand \emph{et~al.}(2013)Le~Grand, Goetz, and Walker]{Le-Grand13}
S.~Le~Grand, A.~W. Goetz and R.~C. Walker, \emph{Comput. Phys. Commun.}, 2013,
  \textbf{184}, 374--380\relax
\mciteBstWouldAddEndPuncttrue
\mciteSetBstMidEndSepPunct{\mcitedefaultmidpunct}
{\mcitedefaultendpunct}{\mcitedefaultseppunct}\relax
\EndOfBibitem
\bibitem[Tompa and Fuxreiter(2008)]{Tompa08}
P.~Tompa and M.~Fuxreiter, \emph{Trends Biochem. Sci.}, 2008, \textbf{33},
  2--8\relax
\mciteBstWouldAddEndPuncttrue
\mciteSetBstMidEndSepPunct{\mcitedefaultmidpunct}
{\mcitedefaultendpunct}{\mcitedefaultseppunct}\relax
\EndOfBibitem
\bibitem[Uversky and Dunker(2012)]{Uversky12}
V.~N. Uversky and A.~K. Dunker, \emph{Anal. Chem.}, 2012, \textbf{84},
  2096--2104\relax
\mciteBstWouldAddEndPuncttrue
\mciteSetBstMidEndSepPunct{\mcitedefaultmidpunct}
{\mcitedefaultendpunct}{\mcitedefaultseppunct}\relax
\EndOfBibitem
\bibitem[Kirschner \emph{et~al.}(2008)Kirschner, Yongye, Tschampel,
  Gonzalez-Outeirino, Daniels, Foley, and Woods]{Kirschner08}
K.~N. Kirschner, A.~B. Yongye, S.~M. Tschampel, J.~Gonzalez-Outeirino, C.~R.
  Daniels, B.~L. Foley and R.~J. Woods, \emph{J. Comput. Chem.}, 2008,
  \textbf{29}, 622--655\relax
\mciteBstWouldAddEndPuncttrue
\mciteSetBstMidEndSepPunct{\mcitedefaultmidpunct}
{\mcitedefaultendpunct}{\mcitedefaultseppunct}\relax
\EndOfBibitem
\bibitem[Dickson \emph{et~al.}(2014)Dickson, Madej, Skjevik, Betz, Teigen,
  Gould, and Walker]{Dickson14}
C.~J. Dickson, B.~D. Madej, A.~A. Skjevik, R.~M. Betz, K.~Teigen, I.~R. Gould
  and R.~C. Walker, \emph{J. Chem. Theory Comput.}, 2014, \textbf{10},
  865--879\relax
\mciteBstWouldAddEndPuncttrue
\mciteSetBstMidEndSepPunct{\mcitedefaultmidpunct}
{\mcitedefaultendpunct}{\mcitedefaultseppunct}\relax
\EndOfBibitem
\bibitem[Helfrich(1978)]{Helfrich78}
W.~Helfrich, \emph{Z. Naturforsch. A}, 1978, \textbf{33}, 305--315\relax
\mciteBstWouldAddEndPuncttrue
\mciteSetBstMidEndSepPunct{\mcitedefaultmidpunct}
{\mcitedefaultendpunct}{\mcitedefaultseppunct}\relax
\EndOfBibitem
\bibitem[Lipowsky and Leibler(1986)]{Lipowsky86}
R.~Lipowsky and S.~Leibler, \emph{Phys. Rev. Lett.}, 1986, \textbf{56},
  2541--2544\relax
\mciteBstWouldAddEndPuncttrue
\mciteSetBstMidEndSepPunct{\mcitedefaultmidpunct}
{\mcitedefaultendpunct}{\mcitedefaultseppunct}\relax
\EndOfBibitem
\bibitem[Steink\"uhler \emph{et~al.}(2019)Steink\"uhler, Rozycki, Alvey,
  Lipowsky, Weikl, Dimova, and Discher]{Steinkuhler19}
J.~Steink\"uhler, B.~Rozycki, C.~Alvey, R.~Lipowsky, T.~R. Weikl, R.~Dimova and
  D.~E. Discher, \emph{J. Cell. Sci.}, 2019, \textbf{132}, jcs216770\relax
\mciteBstWouldAddEndPuncttrue
\mciteSetBstMidEndSepPunct{\mcitedefaultmidpunct}
{\mcitedefaultendpunct}{\mcitedefaultseppunct}\relax
\EndOfBibitem
\bibitem[Weikl(2018)]{Weikl18}
T.~R. Weikl, \emph{Annu. Rev. Phys. Chem.}, 2018, \textbf{69}, 521--539\relax
\mciteBstWouldAddEndPuncttrue
\mciteSetBstMidEndSepPunct{\mcitedefaultmidpunct}
{\mcitedefaultendpunct}{\mcitedefaultseppunct}\relax
\EndOfBibitem
\bibitem[Simson \emph{et~al.}(1998)Simson, Wallraff, Faix, Niewohner, Gerisch,
  and Sackmann]{Simson98}
R.~Simson, E.~Wallraff, J.~Faix, J.~Niewohner, G.~Gerisch and E.~Sackmann,
  \emph{Biophys. J.}, 1998, \textbf{74}, 514--522\relax
\mciteBstWouldAddEndPuncttrue
\mciteSetBstMidEndSepPunct{\mcitedefaultmidpunct}
{\mcitedefaultendpunct}{\mcitedefaultseppunct}\relax
\EndOfBibitem
\bibitem[Popescu \emph{et~al.}(2006)Popescu, Ikeda, Goda, Best-Popescu,
  Laposata, Manley, Dasari, Badizadegan, and Feld]{Popescu06}
G.~Popescu, T.~Ikeda, K.~Goda, C.~A. Best-Popescu, M.~Laposata, S.~Manley,
  R.~R. Dasari, K.~Badizadegan and M.~S. Feld, \emph{Phys. Rev. Lett.}, 2006,
  \textbf{97}, 218101\relax
\mciteBstWouldAddEndPuncttrue
\mciteSetBstMidEndSepPunct{\mcitedefaultmidpunct}
{\mcitedefaultendpunct}{\mcitedefaultseppunct}\relax
\EndOfBibitem
\bibitem[Betz \emph{et~al.}(2009)Betz, Lenz, Joanny, and Sykes]{Betz09}
T.~Betz, M.~Lenz, J.-F. Joanny and C.~Sykes, \emph{Proc. Natl. Acad. Sci. USA},
  2009, \textbf{106}, 15320--15325\relax
\mciteBstWouldAddEndPuncttrue
\mciteSetBstMidEndSepPunct{\mcitedefaultmidpunct}
{\mcitedefaultendpunct}{\mcitedefaultseppunct}\relax
\EndOfBibitem
\bibitem[Nagle(2013)]{Nagle13}
J.~F. Nagle, \emph{Faraday Discuss.}, 2013, \textbf{161}, 11--29\relax
\mciteBstWouldAddEndPuncttrue
\mciteSetBstMidEndSepPunct{\mcitedefaultmidpunct}
{\mcitedefaultendpunct}{\mcitedefaultseppunct}\relax
\EndOfBibitem
\bibitem[Dimova(2014)]{Dimova14}
R.~Dimova, \emph{Adv. Colloid Interface Sci.}, 2014, \textbf{208},
  225--234\relax
\mciteBstWouldAddEndPuncttrue
\mciteSetBstMidEndSepPunct{\mcitedefaultmidpunct}
{\mcitedefaultendpunct}{\mcitedefaultseppunct}\relax
\EndOfBibitem
\bibitem[Xu \emph{et~al.}(2015)Xu, Hu, Lipowsky, and Weikl]{Xu15}
G.-K. Xu, J.~Hu, R.~Lipowsky and T.~R. Weikl, \emph{J. Chem. Phys.}, 2015,
  \textbf{143}, 243136\relax
\mciteBstWouldAddEndPuncttrue
\mciteSetBstMidEndSepPunct{\mcitedefaultmidpunct}
{\mcitedefaultendpunct}{\mcitedefaultseppunct}\relax
\EndOfBibitem
\bibitem[Geyer \emph{et~al.}(2000)Geyer, Gege, and Schmidt]{Geyer00}
A.~Geyer, C.~Gege and R.~R. Schmidt, \emph{Angew. Chem. Int. Edit.}, 2000,
  \textbf{39}, 3246\relax
\mciteBstWouldAddEndPuncttrue
\mciteSetBstMidEndSepPunct{\mcitedefaultmidpunct}
{\mcitedefaultendpunct}{\mcitedefaultseppunct}\relax
\EndOfBibitem
\bibitem[Nodet \emph{et~al.}(2007)Nodet, Poggi, Abergel, Gourmala, Dong, Zhang,
  Mallet, and Bodenhausen]{Nodet07}
G.~Nodet, L.~Poggi, D.~Abergel, C.~Gourmala, D.~Dong, Y.~Zhang, J.-M. Mallet
  and G.~Bodenhausen, \emph{J. Am. Chem. Soc.}, 2007, \textbf{129},
  9080--9085\relax
\mciteBstWouldAddEndPuncttrue
\mciteSetBstMidEndSepPunct{\mcitedefaultmidpunct}
{\mcitedefaultendpunct}{\mcitedefaultseppunct}\relax
\EndOfBibitem
\bibitem[Woods(2005-2018)]{glycamweb}
G.~Woods, 2005-2018\relax
\mciteBstWouldAddEndPuncttrue
\mciteSetBstMidEndSepPunct{\mcitedefaultmidpunct}
{\mcitedefaultendpunct}{\mcitedefaultseppunct}\relax
\EndOfBibitem
\bibitem[Salomon-Ferrer \emph{et~al.}(2013)Salomon-Ferrer, Case, and
  Walker]{Salomon13}
R.~Salomon-Ferrer, D.~A. Case and R.~C. Walker, \emph{Wiley Interdiscip.
  Rev.-Comput. Mol. Sci.}, 2013, \textbf{3}, 198--210\relax
\mciteBstWouldAddEndPuncttrue
\mciteSetBstMidEndSepPunct{\mcitedefaultmidpunct}
{\mcitedefaultendpunct}{\mcitedefaultseppunct}\relax
\EndOfBibitem
\bibitem[Allen and Tildesley(2017)]{Allen17}
M.~P. Allen and D.~J. Tildesley, \emph{Computer simulation of liquids}, Oxford
  University Press, 2017\relax
\mciteBstWouldAddEndPuncttrue
\mciteSetBstMidEndSepPunct{\mcitedefaultmidpunct}
{\mcitedefaultendpunct}{\mcitedefaultseppunct}\relax
\EndOfBibitem
\bibitem[Miyamoto and Kollman(1992)]{Miyamoto92}
S.~Miyamoto and P.~A. Kollman, \emph{J. Comput. Chem.}, 1992, \textbf{13},
  952--962\relax
\mciteBstWouldAddEndPuncttrue
\mciteSetBstMidEndSepPunct{\mcitedefaultmidpunct}
{\mcitedefaultendpunct}{\mcitedefaultseppunct}\relax
\EndOfBibitem
\bibitem[Ryckaert \emph{et~al.}(1977)Ryckaert, Ciccotti, and
  Berendsen]{Ryckaert77}
J.-P. Ryckaert, G.~Ciccotti and H.~J. Berendsen, \emph{Journal of Computational
  Physics}, 1977, \textbf{23}, 327 -- 341\relax
\mciteBstWouldAddEndPuncttrue
\mciteSetBstMidEndSepPunct{\mcitedefaultmidpunct}
{\mcitedefaultendpunct}{\mcitedefaultseppunct}\relax
\EndOfBibitem
\bibitem[Essmann \emph{et~al.}(1995)Essmann, Perera, Berkowitz, Darden, Lee,
  and Pedersen]{Essmann95}
U.~Essmann, L.~Perera, M.~L. Berkowitz, T.~Darden, H.~Lee and L.~G. Pedersen,
  \emph{J. Chem. Phys.}, 1995, \textbf{103}, 8577--8593\relax
\mciteBstWouldAddEndPuncttrue
\mciteSetBstMidEndSepPunct{\mcitedefaultmidpunct}
{\mcitedefaultendpunct}{\mcitedefaultseppunct}\relax
\EndOfBibitem
\bibitem[Darden \emph{et~al.}(1993)Darden, York, and Pedersen]{Darden93}
T.~Darden, D.~York and L.~Pedersen, \emph{J. Chem. Phys.}, 1993, \textbf{98},
  10089--10092\relax
\mciteBstWouldAddEndPuncttrue
\mciteSetBstMidEndSepPunct{\mcitedefaultmidpunct}
{\mcitedefaultendpunct}{\mcitedefaultseppunct}\relax
\EndOfBibitem
\bibitem[Hu \emph{et~al.}(2013)Hu, Lipowsky, and Weikl]{Hu13}
J.~Hu, R.~Lipowsky and T.~R. Weikl, \emph{Proc. Natl. Acad. Sci. USA}, 2013,
  \textbf{110}, 15283--15288\relax
\mciteBstWouldAddEndPuncttrue
\mciteSetBstMidEndSepPunct{\mcitedefaultmidpunct}
{\mcitedefaultendpunct}{\mcitedefaultseppunct}\relax
\EndOfBibitem
\bibitem[Kucerka \emph{et~al.}(2011)Kucerka, Nieh, and Katsaras]{Kucerka11}
N.~Kucerka, M.-P. Nieh and J.~Katsaras, \emph{Biochim. Biophys.
  Acta-Biomembr.}, 2011, \textbf{1808}, 2761--2771\relax
\mciteBstWouldAddEndPuncttrue
\mciteSetBstMidEndSepPunct{\mcitedefaultmidpunct}
{\mcitedefaultendpunct}{\mcitedefaultseppunct}\relax
\EndOfBibitem
\bibitem[Jo \emph{et~al.}(2008)Jo, Kim, Iyer, and Im]{Jo08}
S.~Jo, T.~Kim, V.~G. Iyer and W.~Im, \emph{J Comput Chem}, 2008, \textbf{29},
  1859--1865\relax
\mciteBstWouldAddEndPuncttrue
\mciteSetBstMidEndSepPunct{\mcitedefaultmidpunct}
{\mcitedefaultendpunct}{\mcitedefaultseppunct}\relax
\EndOfBibitem
\bibitem[Berendsen \emph{et~al.}(1984)Berendsen, Postma, van Gunsteren, DiNola,
  and Haak]{Berendsen84}
H.~J. Berendsen, J.~v. Postma, W.~F. van Gunsteren, A.~DiNola and J.~Haak,
  \emph{J. Chem. Phys.}, 1984, \textbf{81}, 3684--3690\relax
\mciteBstWouldAddEndPuncttrue
\mciteSetBstMidEndSepPunct{\mcitedefaultmidpunct}
{\mcitedefaultendpunct}{\mcitedefaultseppunct}\relax
\EndOfBibitem
\bibitem[Anezo \emph{et~al.}(2003)Anezo, de~Vries, Holtje, Tieleman, and
  Marrink]{Anezo03}
C.~Anezo, A.~H. de~Vries, H.~D. Holtje, D.~P. Tieleman and S.~J. Marrink,
  \emph{J. Phys. Chem. B}, 2003, \textbf{107}, 9424--9433\relax
\mciteBstWouldAddEndPuncttrue
\mciteSetBstMidEndSepPunct{\mcitedefaultmidpunct}
{\mcitedefaultendpunct}{\mcitedefaultseppunct}\relax
\EndOfBibitem
\bibitem[Kucerka \emph{et~al.}(2005)Kucerka, Tristram-Nagle, and
  Nagle]{Kucerka05}
N.~Kucerka, S.~Tristram-Nagle and J.~F. Nagle, \emph{J. Membr. Biol.}, 2005,
  \textbf{208}, 193--202\relax
\mciteBstWouldAddEndPuncttrue
\mciteSetBstMidEndSepPunct{\mcitedefaultmidpunct}
{\mcitedefaultendpunct}{\mcitedefaultseppunct}\relax
\EndOfBibitem
\bibitem[Filippov \emph{et~al.}(2003)Filippov, Oradd, and Lindblom]{Filippov03}
A.~Filippov, G.~Oradd and G.~Lindblom, \emph{Langmuir}, 2003, \textbf{19},
  6397--6400\relax
\mciteBstWouldAddEndPuncttrue
\mciteSetBstMidEndSepPunct{\mcitedefaultmidpunct}
{\mcitedefaultendpunct}{\mcitedefaultseppunct}\relax
\EndOfBibitem
\bibitem[Morishita(2000)]{Morishita00}
T.~Morishita, \emph{J. Chem. Phys.}, 2000, \textbf{113}, 2976--2982\relax
\mciteBstWouldAddEndPuncttrue
\mciteSetBstMidEndSepPunct{\mcitedefaultmidpunct}
{\mcitedefaultendpunct}{\mcitedefaultseppunct}\relax
\EndOfBibitem
\end{mcitethebibliography}
\bibliographystyle{rsc} 

\providecommand*{\mcitethebibliography}{\thebibliography}
\csname @ifundefined\endcsname{endmcitethebibliography}
{\let\endmcitethebibliography\endthebibliography}{}

\end{document}